\documentclass[twocolumn, secnumarabic, amssymb, nobibnotes, pre, superscriptaddress]{revtex4}

\usepackage{graphicx} 
\usepackage{amsmath}
\usepackage{float}
\usepackage{color}
\usepackage[english]{babel}
\usepackage{array}
\usepackage[T1]{fontenc}
\usepackage[usenames,dvipsnames]{xcolor}
\usepackage[utf8]{inputenc}
\usepackage{setspace}
\usepackage{bm}
\usepackage{lipsum} 
\usepackage{tikz}
\usetikzlibrary{math}
\usetikzlibrary{tikzmark}
\usetikzlibrary{decorations.pathreplacing}
\usetikzlibrary{arrows.meta}
\usepackage{stmaryrd}

\definecolor{color1}{rgb}{0.122,0.467,0.706}
\definecolor{color2}{rgb}{0.839,0.153,0.157}

\definecolor{darkblue}{rgb}{0,0,0.6}
\definecolor{darkred}{rgb}{0.6,0,0}
\usepackage[colorlinks=true,urlcolor=darkblue,citecolor=darkblue, linkcolor=darkred,hyperfootnotes=false]{hyperref}

\newcommand{\dd}{\text{d}}

\newcommand{\ee}{\boldsymbol{e}}
\newcommand{\mm}{\boldsymbol{m}}
\newcommand{\nn}{\boldsymbol{n}}
\newcommand{\rr}{\boldsymbol{r}}
\newcommand{\uu}{\boldsymbol{u}}

\DeclareMathOperator{\Tr}{Tr}

\makeatletter
\renewcommand*{\fnum@figure}{{\normalfont \small{FIG.}~\thefigure}}
\makeatother

\begin{document}

\title{Noise-induced collective actuation in active solids}

\author{Paul Baconnier}
\affiliation{AMOLF, 1098 XG Amsterdam, The Netherlands.}
\affiliation{UMR CNRS Gulliver 7083, ESPCI Paris, PSL Research University, 75005 Paris, France.}
\author{Vincent Démery}
\affiliation{UMR CNRS Gulliver 7083, ESPCI Paris, PSL Research University, 75005 Paris, France.}
\affiliation{Univ Lyon, ENSL, CNRS, Laboratoire de Physique, F-69342 Lyon, France.}
\author{Olivier Dauchot}
\affiliation{UMR CNRS Gulliver 7083, ESPCI Paris, PSL Research University, 75005 Paris, France.}

\begin{abstract}
Collective actuation describes the spontaneous synchronized oscillations taking place in active solids, when the elasto-active feedback, which generically couples the reorientation of the active forces and the elastic stress, is large enough.
In the absence of noise, collective actuation takes the form of a strong condensation of the dynamics on a specific pair of modes and their generalized harmonics. 
Here we report experiments conducted with centimetric active elastic structures, where collective oscillation takes place along the single lowest energy mode of the system, gapped from the other modes because of the system's geometry.
Combining the numerical and theoretical analysis of an agent-based model, we demonstrate that this form of collective actuation is noise-induced. The effect of the noise is first analyzed in a single-particle toy model that reveals the interplay between the noise and the specific structure of the phase space. We then show that in the continuous limit, any finite amount of noise turns this new form of transition to collective actuation into a bona fide supercritical Hopf bifurcation.
\end{abstract}

\maketitle

\section{Introduction}
Understanding the vibrational excitations of a solid has been a key step in the development of solid-state physics.
At equilibrium, thermal fluctuations distribute uniformly amongst the vibrational modes.
Out of equilibrium, equipartition does not hold, and more selective actuation of the modes may hold.
This is notably the case in the context of active matter.  A simple nonequilibrium-correlated bath actuates an infinitesimal zero mode while simultaneously suppressing fluctuations in higher modes as compared to thermal fluctuations~\cite{woodhouse2018autonomous}.  Furthermore, when the nodes of the elastic network are motile, self-propulsion is able to fully mobilize free-moving mechanisms, even in topologically complex cases~\cite{woodhouse2018autonomous}.

Active matter reveals its richness further when an elasto-active feedback couples the reorientation of the active forces to the strain and stress induced inside the solid by these forces~\cite{henkes2011active, ferrante2013elasticity, baconnier2022selective,zheng2022experimental,baconnier2023discontinuous,hernandez20232}. 
In the presence of zero modes associated with rigid body motion or mechanisms, the dynamics then fully condense onto one of these zero modes~\cite{szabo2006phase,ferrante2013elasticity,zheng2022experimental,hernandez20232}.
In the absence of zero modes, new dynamical behaviors emerge in the form of spontaneous collective and synchronous oscillations~\cite{henkes2011active, baconnier2022selective, baconnier2023discontinuous}, strikingly resembling the oscillations observed in dense living active systems \cite{henkes2011active, serra2012mechanical, deforet2014emergence, barton2017active, petrolli2019confinement, peyret2019sustained, liu2021viscoelastic, xu2023autonomous}. Taking advantage of a mechanical model system made of springs and self-propelling toys, called Hexbugs, it was recently shown that such oscillations result from the selection of a pair of extended modes, with orthogonal polarization. These modes are not necessarily the lowest energy ones~\cite{baconnier2022selective,baconnier2023discontinuous}. The selection mechanism is rooted in the elasto-active feedback coupling the structure's deformations and the reorientation of the active forces. 

The effect of noise on the dynamics reported above remains largely unexplored. In the case where there are multiple zero modes associated with solid body motions and/or mechanisms, the dynamics can be mapped onto an effective free energy and the dynamics quasi-statically follow the zero mode of minimal free energy~\cite{hernandez20232}. In the absence of zero modes, the selective and collective actuation is robust with respect to the addition of a finite amount of white noise~\cite{baconnier2022selective}.  
A more unexpected effect of noise is the one that we uncover in the present work: the onset of a noise-induced collective actuation (NICA) regime, which takes place along the single lowest energy mode.  Starting from the experimental observation of a regular oscillation in an active elastic structure, the lowest energy mode of which is gapped for geometrical reasons, we first demonstrate, using numerical simulations, that the oscillations only take place in the presence of noise on the orientational dynamics of the active forces.
We then discuss the emergence of these oscillations in two opposite asymptotic limits: that of a single active particle confined in a one-dimensional harmonic potential and that of the continuum limit, using a coarse-grained description of the dynamics. While the study of the single particle provides evidence that the oscillations take their roots in the specific structure of the phase space, the coarse-grained analysis firmly demonstrates that any finite amount of noise makes this new form of transition to collective actuation a bona fide supercritical Hopf bifurcation.
\section{Collective actuation of Active Ladders}

\subsection{Theoretical background}
The collective actuation of active elastic structures has been described in detail in~\cite{baconnier2022selective}. 
Let us nevertheless start, for completeness and self-consistency, with a brief overview of the existing results about collective actuation in the absence of noise. 
Consider a mechanically stable elastic structure composed of active nodes connected by springs of stiffness $k$. 
At each node $i$, an active unit is exerting an active force of amplitude $F_0$ in the direction of its polarity vector, $\hat \nn_i$. The latter is free to rotate, and reorients towards the displacement of the node, dictated by the sum of the forces acting on it. This nonlinear feedback between deformations and polarizations is characterized by two length scales: (i) the elastic length $l_e = F_0/k$, the typical elastic deformation caused by the active forces, and (ii) the self-alignment length $l_a$, the characteristic distance on which a node must be displaced to reorient the active force located at that node. The ratio $\Pi = l_e/l_a$, which we refer to as the elasto-active coupling, controls the transition from a disordered frozen state, with all active forces pointing in random directions, to an oscillating state, where all the nodes oscillate synchronously around their reference position. In this regime, the dynamics condensate on a pair of modes and their generalized harmonics. The selected modes are not necessarily the lowest energy ones; they should also be maximally extended and their polarization should be locally orthogonal. Within the harmonic approximation, i.e., linear elasticity, the dynamics are well described by
\begin{subequations} \label{eq:N_harmonic_approx}
\begin{align}
 \dot{\boldsymbol{u}}_i &= \Pi \hat\nn_i - \mathbb{M}_{ij} \boldsymbol{u}_j, \label{eq1:N_harmonic_approx} \\
 \dot{\boldsymbol{n}}_i &= ( \hat\nn_i \times \dot{\boldsymbol{u}}_i ) \times \hat\nn_i + \sqrt{2D} \xi_i \hat\nn_i^{\perp}, \label{eq2:N_harmonic_approx}
\end{align}
\end{subequations}
where $\boldsymbol{u}_i$ is the displacement of node $i$ with respect to the reference configuration, and $\mathbb{M}$ is the dynamical matrix, whose eigenvectors are the normal, or vibrational, modes of the elastic structure. Equation~(\ref{eq2:N_harmonic_approx}) governs the reorientation of the particles, and contains the nonlinear self-alignment term, and a noise term of amplitude $D$, with correlations $\langle \xi_i(t) \xi_j (t') \rangle = \delta_{ij} \delta(t - t')$. Note that the above equations are made dimensionless, using $\gamma/k$ and $l_a$ as units of time and length, respectively, where $\gamma$ is an effective friction coefficient.

Reducing the dynamics to the two selected modes allows us to map the $N$-body dynamics onto that of a single self-aligning polar particle in a harmonic potential. In the simplest case, when the two modes are degenerated, the oscillating dynamics correspond to simple circular orbits~\cite{dauchot2019dynamics}.
This dynamics was experimentally reported in the case of a triangular lattice pinned at its edges, in which the collective actuation takes the form of a \textit{synchronized chiral oscillation} of the nodes around their mechanical equilibrium position~\cite{baconnier2022selective}. The generalization to two nondegenerated modes was thoroughly investigated in~\cite{damascena2022coexisting} and was experimentally observed when the central node of the same triangular lattice is pinned in both translation and rotation, leading to a \textit{global alternating rotation}~\cite{baconnier2023discontinuous}.
In all cases, it is clearly established that no oscillations can take place when reducing the dynamics to a single mode. 
In the following, we shall see that the latter statement holds only in the absence of noise.

\subsection{Experiments}
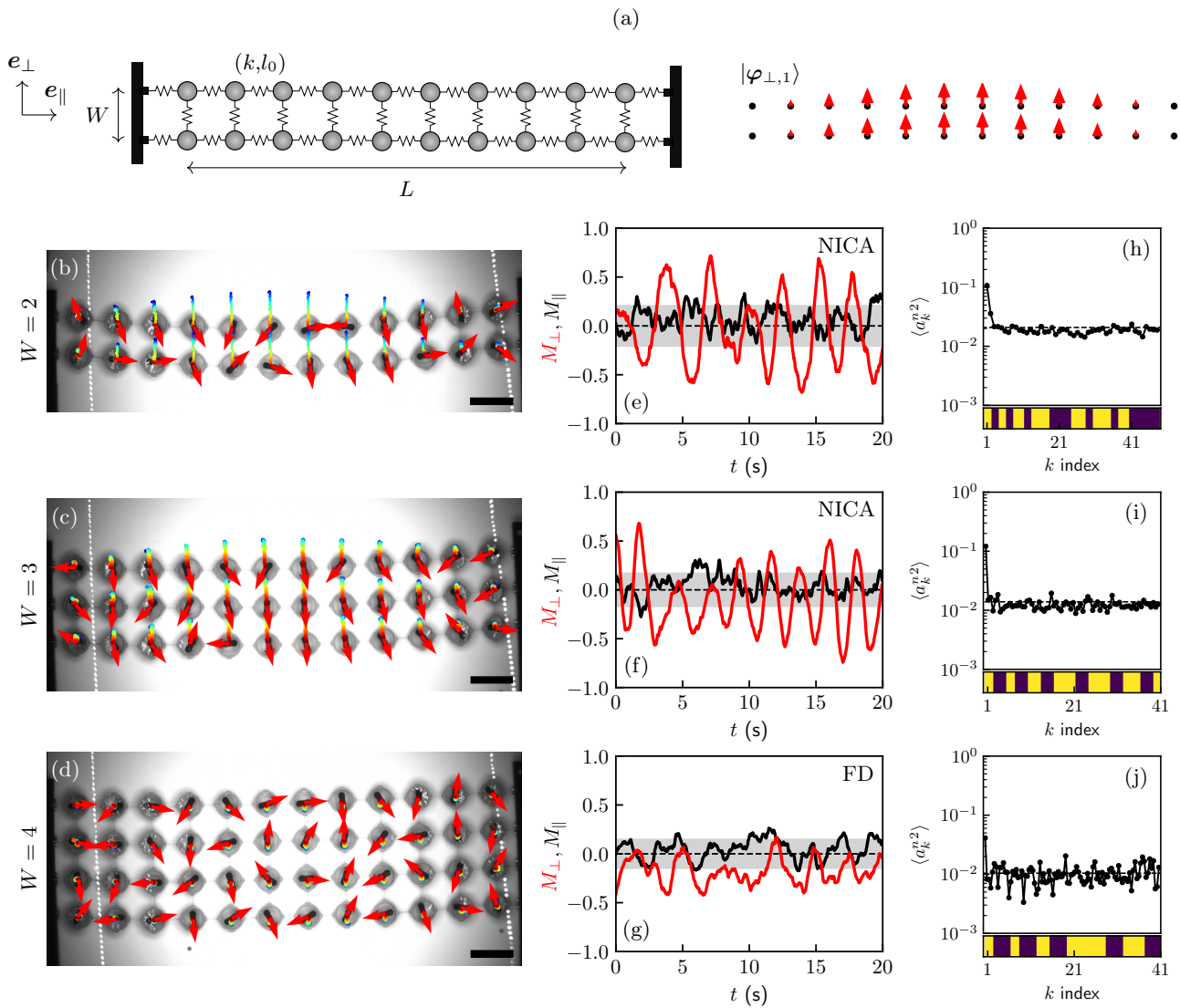
\begin{figure*}[t!]
\hspace*{0.4cm}
 \begin{tikzpicture}
 
 \node[rotate=-90] at (9.8,7.0) {\includegraphics[height=12.0cm]{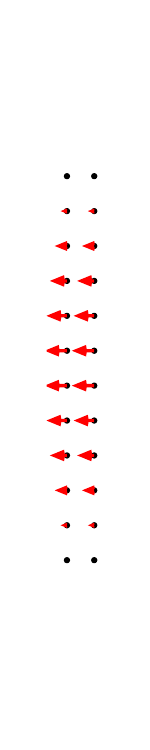}};
 \node[rotate=0] at (7.1,7.6) {\small $| \boldsymbol{\varphi}_{\perp,1} \rangle$};
 
 \node[rotate=0] at (1.8,7.0) {\includegraphics[width=8.0cm]{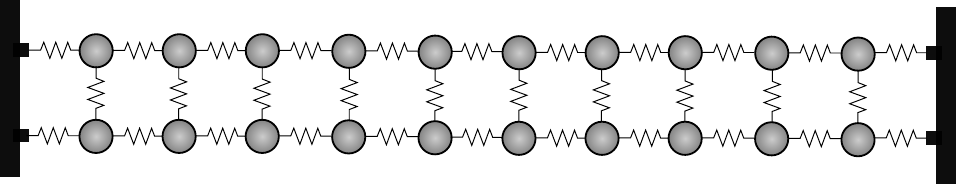}};
 
 \draw[<->] (-1.4,6.23) -- (5.0,6.23);
 \node[rotate=0] at (1.8,5.93) {\small $L$};
 
 \draw[<->] (-2.4,7.4) -- (-2.4,6.6);
 \node[rotate=0] at (-2.7,7.0) {\small $W$};
 
 \node[rotate=0] at (-0.33,7.75) {\small ($k$,$l_0$)};

 \draw[->] (-3.8,7.0) -- (-3.3,7.0) node[above=0.5] {$\boldsymbol{e}_{\parallel}$};
 \draw[->] (-3.8,7.0) -- (-3.8,7.5) node[above=0.5] {$\boldsymbol{e}_{\perp}$};
 
 \node[anchor=east, rotate=0] at (3.6,3.85) {\includegraphics[width=6.92cm]{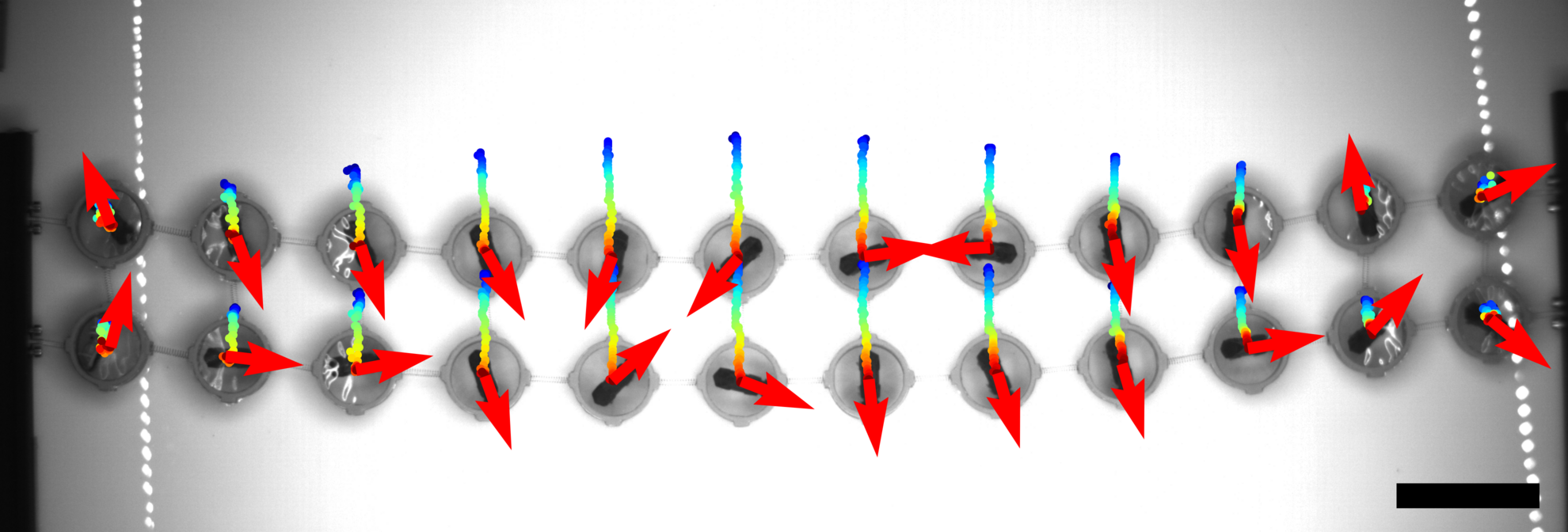}};
 \node[rotate=0] at (6.3,3.6) {\includegraphics[height=3.8cm]{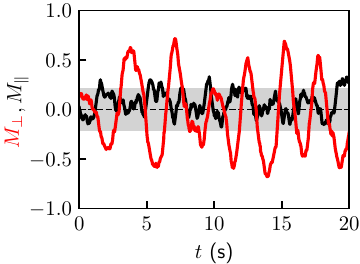}};
 
 \node[anchor=east, rotate=0] at (3.6,0.0) {\includegraphics[width=6.92cm]{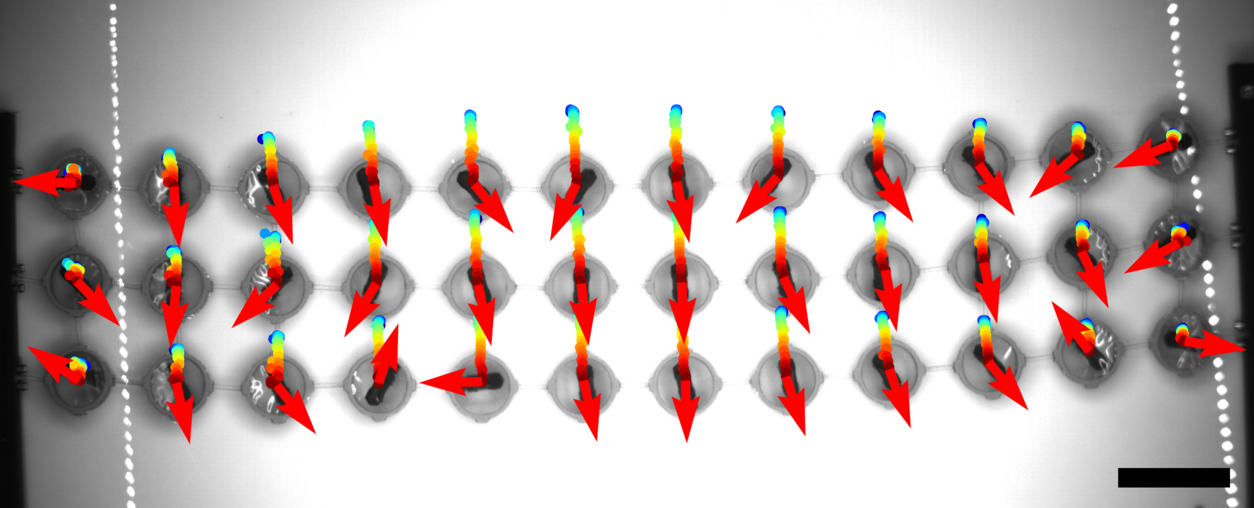}};
 \node[rotate=0] at (6.3,-0.25) {\includegraphics[height=3.8cm]{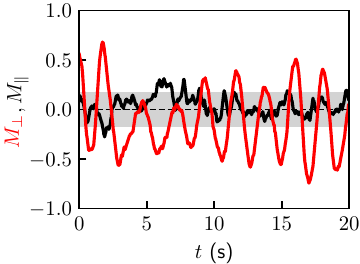}};
 
 \node[anchor=east, rotate=0] at (3.6,-3.85) {\includegraphics[width=6.92cm]{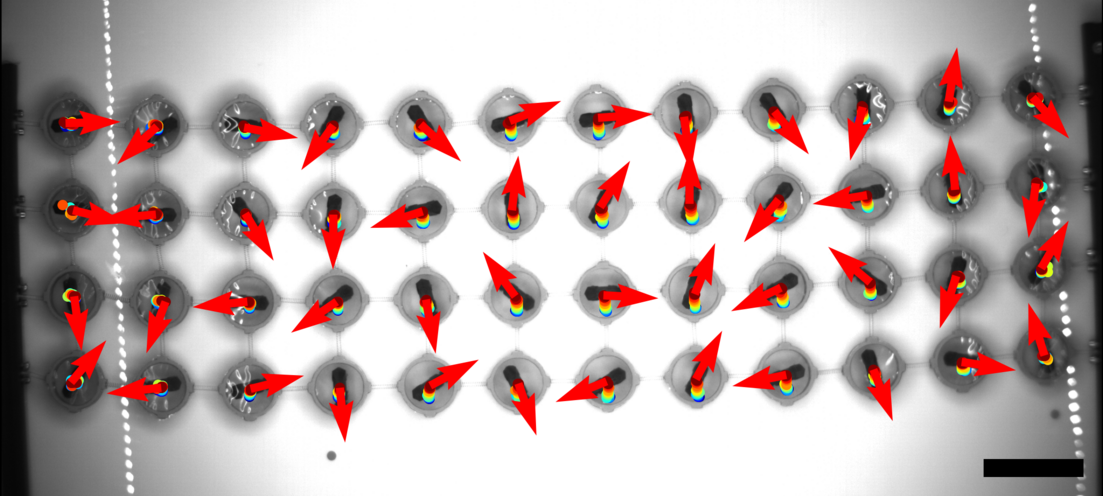}};
 \node[rotate=0] at (6.3,-4.1) {\includegraphics[height=3.8cm]{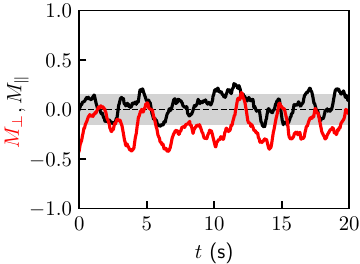}};
 
 \node[rotate=0] at (11.0,3.6) {\includegraphics[height=3.8cm]{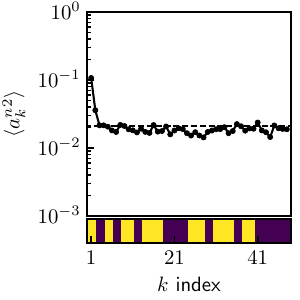}};
 \node[rotate=0] at (11.0,-0.25) {\includegraphics[height=3.8cm]{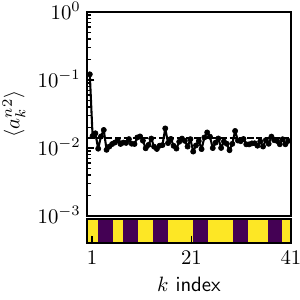}};
 \node[rotate=0] at (11.0,-4.1) {\includegraphics[height=3.8cm]{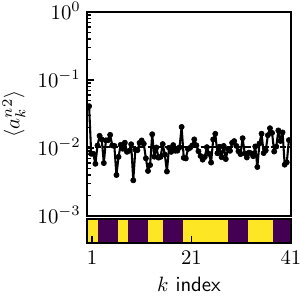}};
 
 \node[rotate=0] at (-3.18,4.75) {\small \color{white} (b)};
 \node[rotate=0] at (-3.18,1.12) {\small \color{white} (c)};
 \node[rotate=0] at (-3.17,-2.57) {\small \color{white} (d)};
 
 \node[rotate=90] at (-3.7,3.85) {\small $W = 2$};
 \node[rotate=90] at (-3.7,0.0) {\small $W = 3$};
 \node[rotate=90] at (-3.7,-3.85) {\small $W = 4$};
 
 \node[rotate=0] at (5.15,2.80) {\small (e)};
 \node[rotate=0] at (5.15,-1.05) {\small (f)};
 \node[rotate=0] at (5.15,-4.9) {\small (g)};
 
 \node[rotate=0] at (8.21,5.1) {\small NICA};
 \node[rotate=0] at (8.21,1.25) {\small NICA};
 \node[rotate=0] at (8.38,-2.6) {\small FD};
 
 \node[rotate=0] at (12.43,5.03) {\small (h)};
 \node[rotate=0] at (12.43,1.18) {\small (i)};
 \node[rotate=0] at (12.43,-2.67) {\small (j)};
 
 \node[rotate=0] at (5.0,8.4) {\small (a)};
  
 \end{tikzpicture}
 \vspace*{-0.5cm}
\caption{\small{\textbf{Experimental observation of collective actuation in active ladders.} (a) Schematic representation of the active ladder architecture (left), and its lowest-energy mode $| \boldsymbol{\varphi}_{\perp,1} \rangle$ (right). (b-d) Snapshot of the dynamics: each row corresponds to a different width $W$; red arrows: polarities $\boldsymbol{\hat{n}}_i$; trajectories color coded from blue to red with increasing time; scale bars: $10$ cm. (e-g) Transverse $M_{\perp}$ (red) and longitudinal $M_{\parallel}$ (black) polarizations as a function of time, corresponding to the dynamics reported in panels (b-d). The gray areas indicate the polarization range $\big[-1/\sqrt{N}, 1/\sqrt{N} \big]$ expected for a purely random dynamics of the polarities. (h-j) Mean squared projections of the polarity field on the normal modes as a function of the modes' index $k$. The dashed horizontal black line indicates equirepartition of the active force. The bottom color bar codes for the modes' geometry (yellow: transverse; blue: longitudinal).}}
\label{fig:zero_gravity}
\end{figure*}
The model active solids considered here consists of elastic structures composed of $N$ active units connected by coil springs of rest length $l_0$ (see~\cite{baconnier2022selective} for details). 
Each active unit is made of a \textit{Hexbug}\textcopyright, a centimetric battery-powered running robot, embedded in a 3D printed annulus (height $1.4$ cm; internal radius $2.5$ cm, $3$ mm thick). Each node has a well-defined reference position set by mechanical equilibrium of the spring network.

We focus on square lattices with rectangular shapes, composed of $L$ (resp. $W$) active units along the long (resp. short) direction of the structure (Figs.~\ref{fig:zero_gravity}-b to d). The left and right ends of the lattices are held fixed so that the extension of the springs in the structure along the longitudinal direction is $\alpha = l_\textrm{eq}/l_0 \simeq 1.28$ in the reference configuration. Doing so, the structure intentionally has a stiff longitudinal direction and a soft transversal one.
Once doped with active units, the dynamics observed in such \textit{active ladders} depends on their width $W$. 
For $W=4$, the system is disordered; the polarities of the active units are diffusing and the displacements of the nodes are small.  For $W=3$ and $W=2$, a global oscillation of the structure, perpendicular to its longitudinal axis takes place (see Supplemental Material movies 1-3).
More quantitatively, we define the longitudinal (resp. transverse) polarizations $M_{\parallel/\perp}(t) = (1/N) \sum_i \hat\nn_i (t) \cdot \boldsymbol{e}_{\parallel/\perp}$, where the vectors $\boldsymbol{e}_{\parallel}$ and $\boldsymbol{e}_{\perp}$ are defined in Fig. \ref{fig:zero_gravity}-a. 
As can be seen from Figs.~\ref{fig:zero_gravity}-e to g, this oscillation is driven by the collective alignment of the polarities of the active units along the transverse direction, while the longitudinal polarization remains within the range expected for random independent orientations. The alignment is not perfect, however, because of the presence of a significant angular noise in the dynamics of the polarities. Within the framework of Eqs. (\ref{eq:N_harmonic_approx}), we could evaluate $D \approx 10^{-1}$ (see Supplementary Information of \cite{baconnier2022selective}, Sec. 2.2). We also note that the oscillation frequency of the ladder of width $W=3$ is larger than that of width $W=2$.
We finally measure $a_k^n = \langle \boldsymbol{\varphi}_k | \hat\nn \rangle/\sqrt{N}$, the normalized projections of the active forces on the normal modes of the idealized spring networks (see Fig.~\ref{fig:zero_gravity}-a and below). As indicated by the square projection averaged over time (Figs.~\ref{fig:zero_gravity}-h to j), the active forces condensate on one mode, the first transverse, lowest-energy one, shown in the right of Fig.~\ref{fig:zero_gravity}-a (see Supplemental Material for a representation of all the modes and their associated eigenvalues or energies \cite{supplentary_information}).

The collective actuation dynamics reported above is therefore very different from the previously reported \textit{synchronized chiral oscillations}~\cite{baconnier2022selective} and \textit{global alternating rotations}~\cite{baconnier2023discontinuous}, the distinctive signature of which is to require the activation of two modes, selected by their specific geometries. While numerical simulations and analytical studies demonstrated that the latter collective actuations develop in the absence of noise, we shall see below that noise is a necessary ingredient for the type of collective actuation reported here to take place.
 
\subsection{Numerical simulations}

We consider Eqs. (\ref{eq:N_harmonic_approx}) for the dynamics of $N=24$ overdamped self-aligning active particles, located at the nodes of the $W=2$ idealized spring network. Because of the elongated geometry of the structure, the first transverse mode $| \boldsymbol{\varphi}_{1} \rangle = | \boldsymbol{\varphi}_{\perp,1} \rangle$ (Fig. \ref{fig:zero_gravity}-a, right) has an energy $\omega_{1}^{2} \simeq 0.01$, significantly lower than the second lowest-energy mode, for which $\omega_{2}^{2} \propto 0.06$.
\begin{figure}[t!]
\centering
\hspace*{0.05cm}
 \begin{tikzpicture}
 
 \node[] at (0.0,-0.57) {\includegraphics[height=4.2cm]{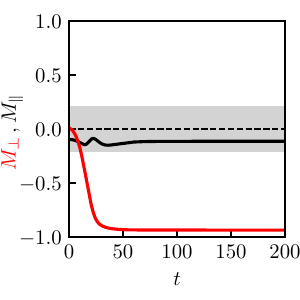}};
 \node[] at (0.0,3.93) {\includegraphics[height=4.2cm]{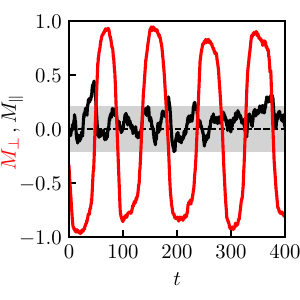}};
 \node[] at (0.0,8.43) {\includegraphics[height=4.2cm]{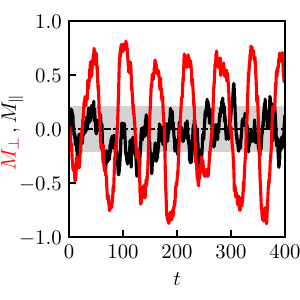}};
 
 \node[] at (4.2,-0.6) {\includegraphics[height=4.0cm]{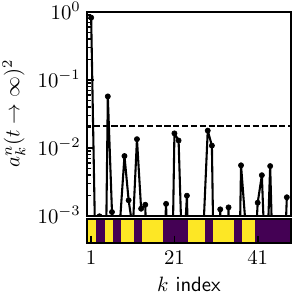}};
 \node[] at (4.2,3.9) {\includegraphics[height=4.0cm]{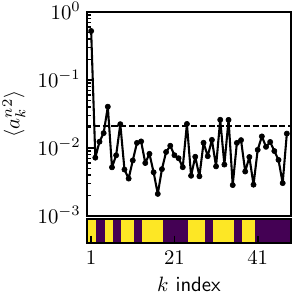}};
 \node[] at (4.2,8.4) {\includegraphics[height=4.0cm]{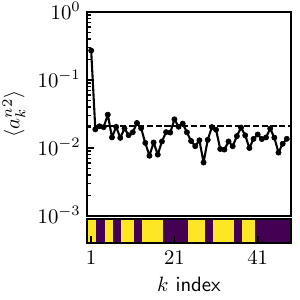}};
 
 \node[anchor=east] at (0.7,10.6) {\small (a)};
 \node[anchor=east] at (0.7,6.1) {\small (b)};
 \node[anchor=east] at (0.7,1.6) {\small (c)};
 
 \node[anchor=east] at (5.1,10.6) {\small (d)};
 \node[anchor=east] at (5.1,6.1) {\small (e)};
 \node[anchor=east] at (5.1,1.6) {\small (f)};

 \end{tikzpicture}
 \vspace*{-0.7cm}
\caption{\small{\textbf{Numerical investigation of the effect of noise on the collective actuation of model active ladders}. Simulations of a model ladder of width $W = 2$, with $N = 24$ nodes and fixed $\Pi/\omega_{1}^{2} = 4.17$, for three different noise amplitudes: $D = 0$ (bottom), $D = 10^{-3}$ (middle), $D = 6.10^{-3}$ (top). (a-c) Transverse $M_{\perp}$ (red) and longitudinal $M_{\parallel}$ (black) polarizations as a function of time.  
(d-f) Mean squared projections of the polarity field on the normal modes as a function of the modes' index $k$. The dashed horizontal black line indicates equirepartition of the active force. The bottom color bar codes for the modes' geometry (yellow: transverse; blue: longitudinal).}}
\label{fig:simulation_noise}
\end{figure}
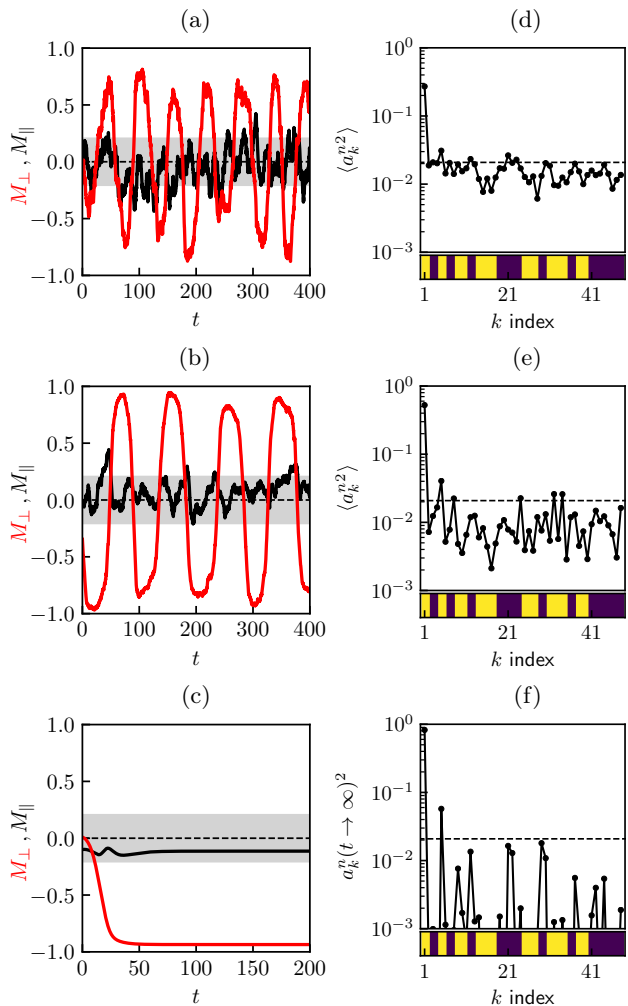
Note that in the experimental system, the width of the ladder controls the transverse stiffness, which results from the finite size of the nodes and the bending and shear moduli of the real springs. This is not the case in the idealized lattice, with purely central forces acting on point-like nodes, for which the transverse stiffness is independent of $W$ (see Supplemental Material \cite{supplentary_information}). In the present simplified model, the stiffness of the lattice is thus controlled utterly by that of the springs, hence by the value of $\Pi$: the larger $\Pi$, the softer the network. We keep $W=2$ fixed, unless stated otherwise. \\
\begin{figure*}[t!]
\centering
 \begin{tikzpicture}

 \node[] at (-0.2,0.0) {\includegraphics[height=5.4cm]{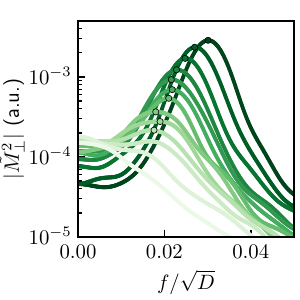}};
 \draw[->] (0.7,-1.1) -- (1.53,1.43);
 \node[] at (1.78,1.68) {\small $\Pi \nearrow$};
 
 \node[] at (-0.2,-5.75) {\includegraphics[height=5.4cm]{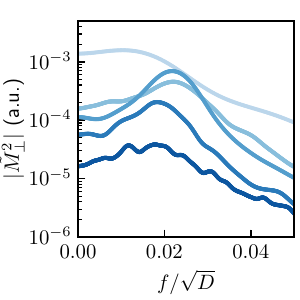}};
 \draw[<-] (0.5,-7.0) -- (1.63,-4.02);
 \node[] at (1.1,-6.8) {\small $D \nearrow$};
 
  \node[] at (5.3,0.0) {\includegraphics[height=5.2cm]{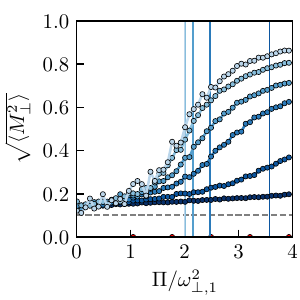}};
 \node[] at (5.3,-5.75) {\includegraphics[height=5.4cm]{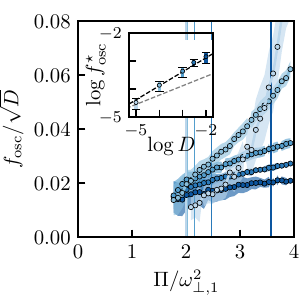}};
 
 \node[rotate=33] at (5.27,-4.45) {\footnotesize $1/2$};
 \node[rotate=21] at (6.12,-4.75) {\footnotesize \color{gray} $1/3$};
  
 \node[] at (10.95,0.0) {\includegraphics[height=5.2cm]{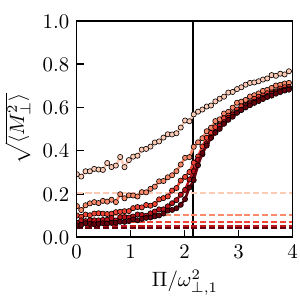}};
 \node[] at (10.95,-5.75) {\includegraphics[height=5.4cm]{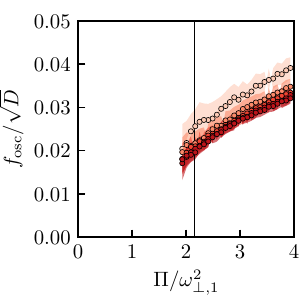}};

 \node[] at (5.2,0.5) {\small $D \nearrow$};
 \draw[<-] (5.8,-1.15) -- (5.25,0.15);
 
 \node[] at (12.55,-0.45) {\small $N \nearrow$};
 \draw[->] (11.1,0.5) -- (12.1,-0.28);
 
 \node[] at (0.3,2.7) {\small (a)};
 \node[] at (5.75,2.7) {\small (c)};
 \node[] at (11.45,2.7) {\small (e)};
 
 \node[] at (0.25,-2.94) {\small (b)};
 \node[] at (5.75,-2.94) {\small (d)};
 \node[] at (11.4,-2.94) {\small (f)};
 
 \end{tikzpicture}
 \vspace*{-0.35cm}
\caption{\small{\textbf{Numerical investigation of the transition towards the collective actuation of model active ladders}. (a-d) Simulations of a model ladder of width $W = 2$, with $N = 24$ nodes. (a, b) Power spectrums obtained from the Fourier transform of $M_{\perp}(t)$; (a) for fixed fixed $D = 10^{-3}$, color coded from light to dark green as $\Pi$ increases, with $\Pi/\omega_{1}^{2} \in \left[ 0.89, 1.21, 1.45, 1.61, 1.77, 1.85, 1.93, 2.01, 2.17, 2.33, 2.49, 2.81, 3.13 \right]$; (b) for fixed $\Pi/\omega_{\perp,1}^{2} = 2$, color coded from light to dark blue as noise increases, with $D \in \left[10^{-5}, 10^{-4}, 10^{-3}, 3.10^{-3}, 10^{-2}\right]$;  (c) Root mean squared transverse polarization $\sqrt{\langle M_{\perp}^2 \rangle}$ as a function of $\Pi$, obtained for different noise amplitudes $D \in \left[10^{-5}, 10^{-4}, 10^{-3}, 3.10^{-3}, 10^{-2}, 3.10^{-2}\right]$, color coded from light to dark blue as noise increases; the dashed horizontal lines indicate the expected value $1/\sqrt{2N}$ for a collection of random independent orientations; the solid vertical blue lines indicate the values $\Pi_c = 2( \omega_{1}^{2} + D)$; (d) Frequency of oscillation as a function of $\Pi$, as obtained from the largest peak of $M_{\perp}(t)$'s Fourier transform, the same conventions as in (c); the shaded areas represent the width of the Gaussian fitted at the vicinity of of largest peak of the $M_{\perp}(t)$ Fourier transform. Inset: log-log plot of the frequency of oscillation at onset, $f_\textrm{osc}^{\star}$ (defined in the text), as a function of noise $D$; the dashed black (resp. gray) line indicates slope $1/2$ (resp. $1/3$). (e, f) Same as (c) and (d) for systems of increasing number of nodes $N \in [6, 24, 54, 96, 150]$ at fixed aspect ration $W/L$ and physical size; $D = 10^{-3}$; the solid black line indicates the value $\Pi_c = 2( \omega_{1}^{2} + D)$.}}
\label{fig:simulation_hopf}
\end{figure*}

We simulate Eqs. (\ref{eq:N_harmonic_approx}) with an Euler-Maruyama method, and a fixed time step $\delta t = 10^{-2}$.  The active units are initiated with positions in the reference configuration of the lattice and with random initial orientations.
When $\Pi < \omega_{1}^{2}$, the system is always disordered (not shown here): in the absence of noise, it is frozen in a mechanical equilibrium configuration set by the random initial orientation of the active units; at finite noise, the orientations diffuse. For $\omega_{1}^{2} < \Pi = 0.05 < \omega_{2}^{2}$, the observed dynamics depend on the presence of noise, as clearly evidenced in Figs.~\ref{fig:simulation_noise}.
For any finite amount of noise, $D > 0$, oscillations emerge along the transverse direction (Figs.~\ref{fig:simulation_noise}-a and b). 
These oscillations resemble those observed experimentally, with a particularly large amplitude along the lowest-energy mode $| \boldsymbol{\varphi}_{\perp,1} \rangle$, while the rest of the spectrum remains essentially flat (Figs.~\ref{fig:simulation_noise}-d and e).
Most importantly, these oscillations do not take place in the absence of noise.
For $D = 0$, the system is \textit{frozen-polarized} (Figs.~\ref{fig:simulation_noise}-c and f): a static state is reached, with the active units pointing in the direction of $| \boldsymbol{\varphi}_{\perp,1} \rangle$, leading to a large and constant value of $\left| M_{\perp} \right|$ while $M_{\parallel}$ remains small.

The role of the noise is further characterized by examining the Fourier spectrum of $M_{\perp}$ as the elasto-active coupling $\Pi$ increases from below $\omega_{1}^{2}$, for different noise amplitudes $D$, and different system sizes (Fig.~\ref{fig:simulation_hopf}). For a given intermediate noise level, the Fourier spectrum of $M_{\perp}$ exhibits a distinctive peak, the amplitude of which increases with $\Pi$ (Fig.~\ref{fig:simulation_hopf}-a).
The same trend is observed when measuring the mean squared transverse polarization, $\langle M_{\perp}^{2} \rangle$, which grows from the small value expected for independent random orientations and saturates close to $1$ for low enough noise (Fig.~\ref{fig:simulation_hopf}-c). 
Altogether, the polarization periodically reverses its orientation with a characteristic frequency $f_\textrm{osc}$, captured by the location of the peak, which increases with $\Pi$ from a finite value $f_\textrm{osc}^{\star}$ at the onset. 
The dependence on the noise amplitude is more subtle.  On one hand, the smaller the noise, the sharper the onset of the oscillation regime (Fig.~\ref{fig:simulation_hopf}-c). On the other hand, at low noise, the peak is absorbed by a strong low-frequency component of the Fourier spectrum (Fig.~\ref{fig:simulation_hopf}-b):  the polarization dynamics is dominated by large amplitude stochastic reversals (Fig.~\ref{fig:simulation_hopf}-c). Finally, very large noise randomizes the dynamics, decreasing both the amplitude of the peak in the Fourier spectrum (Fig.~\ref{fig:simulation_hopf}-b)  and the amplitude of $\langle M_{\perp}^{2} \rangle$ (Fig.~\ref{fig:simulation_hopf}-c). In the range of noise where the oscillations are well captured, we note a square root dependence of the oscillation frequency at onset, $f_\textrm{osc}^{\star}$, with the noise amplitude.

Finally, we perform simulations with increasing values of $N$, while keeping constant the physical size $L$ and the aspect ratio $L/W$. This amounts to decreasing the lattice spacing, or, equivalently, to rescaling the eigenvalues of the dynamical matrix so that the lowest-energy mode keeps the same energy. The noise, $D = 10^{-3}$, is kept constant. The onset of oscillations is sharper with increasing system size (Fig. \ref{fig:simulation_hopf}-e), while the oscillating frequency, extracted from an increasingly sharper peak (not shown here) is essentially independent of $N$ (Fig.~\ref{fig:simulation_hopf}-f).

\subsection{Summary}
Performing experiments with elongated active elastic structures, we observe a transition from a disordered dynamics to a global oscillation of the structure, when the aspect ratio of the structure is large enough. The oscillation condensates on a single transverse mode, which is also the lowest energy one. Simulating an idealized version of the experimental structures in the presence of a finite amount of noise, the same transition is observed when the elasto-active coupling $\Pi$ exceeds the energy of the lowest-energy mode, while remaining smaller than that of the second one. Most importantly, the transition observed numerically is suppressed in the absence of noise. The present collective actuation is thus noise-induced, in sharp contrast with the collective actuation regimes reported before. This, together with the condensation of the dynamics on a single mode, underlines its different nature, hence the name noise-induced collective actuation (NICA).

\section{Noise-induced oscillations in a single particle model}

\subsection{From $N$ to $1$ particle}

In this section, we discuss the onset of the oscillating dynamics by analyzing the stochastic dynamics of a single-particle toy model. 
Our starting point is the Langevin dynamics described by Eqs.~(\ref{eq:N_harmonic_approx}).
In the absence of noise, one sees that any configuration of the polarity field $\{ \hat\nn_i \}$ is a fixed point of the dynamics defined by ($\{\hat\nn_i \},\{ \uu_i = \Pi \mathbb{M}_{ij}^{-1} \hat\nn_j\}$). 
The linear stability analysis of any of these fixed points demonstrates that the linearized dynamics has a zero eigenvalue for all values of $\Pi$. The destabilization threshold $\Pi_c (\{\hat \nn_i \})$, below which the considered fixed point is marginal, and above which it is linearly unstable, depends on the orientation of the active units. 
The first $2^N$ configurations to destabilize are those where the active forces point orthogonally to the lowest energy mode. This destabilization takes place for $\Pi = \Pi_c^{\text{min}} = \omega_{\text{min}}^{2}$, where $\omega_{\text{min}}^{2}$ is the smallest eigenvalue of the dynamical matrix $\mathbb{M}$. The eigenvector associated with this eigenvalue describes a linearized dynamics, where the active units and the displacements point perpendicularly to the lowest energy mode~\cite{baconnier2022selective}. 
In the present case of active ladders, the lowest energy mode is the transverse mode $| \boldsymbol{\varphi}_{\text{min}} \rangle = | \boldsymbol{\varphi}_{\perp,1} \rangle$ illustrated in Fig.~\ref{fig:zero_gravity}-a. Hence all the configurations where the active forces point in the longitudinal direction become unstable when $\Pi > \omega_{1}^{2}$. The active units orient along the transverse direction and the ladder adopts a bent shape, the amplitude of which is set by $\Pi$. This global deformation of the ladder spontaneously breaks the transversal symmetry, and separates the high-dimensional set of fixed points into two subsets of marginally stable fixed points, respectively polarized along and opposite to $| \boldsymbol{\varphi}_{\text{min}} \rangle = | \boldsymbol{\varphi}_{\perp,1} \rangle$.  

The dynamics thus effectively reduces to a one-dimensional dynamics along the softest mode, and, in the presence of noise, one would expect to observe stochastic inversions between these two subsets rather than an oscillation with a typical frequency. The situation is however not as simple. First, the dynamics is fully out of equilibrium and not thermally activated. Second, the two subsets are composed of marginally stable fixed points and should not be seen as basins of attraction.
The physics associated with this specific organization of the phase space is already well captured by a single particle model, which we describe below. 

\subsection{A single particle confined to 1d}
In the case of the previously reported collective actuation~\cite{baconnier2022selective}, when the dynamics condensates on two modes, the natural single particle model is that of a self-aligning particle in a two-dimensional harmonic potential, with stiffnesses given by the energies of the two selected modes. When the two modes are degenerated, with the same energy $\omega_{0}^{2}$, the harmonic potential is rotationally invariant, thus all fixed points are equivalent and become unstable for $\Pi > \omega_{0}^{2}$, leading to chiral oscillations in the form of circular orbits \cite{dauchot2019dynamics}. When the two modes have different energies $\omega_{1}^{2} < \omega_{2}^{2}$, invariance by rotation is broken, the fixed points destabilize progressively in the range $\omega_{1}^{2} < \Pi < \omega_{2}^{2}$, and the chiral oscillations transform into a family of elliptic regimes elongated along the softer direction of the potential~\cite{damascena2022coexisting}. 

The situation one wants to consider here is the limiting case where $\omega_{1}^{2} \ll \omega_{2}^{2}$ and $\Pi$ remains comparable with the energy of the lowest energy mode. It corresponds to that of a single particle moving along a rigid rail, with the elastic restoring force being parallel to the rail's axis, while its polarity can still take all possible orientations.
$u$ denotes the displacement along the soft direction (say, of axis $\hat\ee_{x}$), and $\theta\in[-\pi, \pi]$ denotes the orientation of $\hat\nn$ with respect to $\hat\ee_{x}$. 
In such a setting, Eqs.~(\ref{eq:N_harmonic_approx}) read:
\begin{subequations} \label{eq:SP_rail_noise}
\begin{align}
\dot u &= \Pi\cos\theta - \omega_{1}^{2} u,\\
\dot \theta &= -\sin\theta[\Pi \cos\theta - \omega_{1}^{2} u] + \sqrt{2D}\xi.
\end{align}
\end{subequations}
The specificity of this model is that the amplitude of the active force effectively depends on its orientation. 
In this simple geometry, one can further rescale time $t\rightarrow \tilde t = \omega_1^2 t$ and displacements $u \rightarrow v= \omega_1^2 u / \Pi$ to reduce to two the number of independent parameters and have them both in the equation for the angular dynamics:
\begin{subequations} \label{eq:dotvth}
\begin{align}
\dot v &= \cos\theta -  v, \label{eq:dotv} \\
\dot \theta &= - \frac{\Pi}{\omega_1^2} \sin\theta [ \cos\theta - v] + \sqrt{\frac{2D}{\omega_1^2}}\xi. \label{eq:dotth_v}
\end{align}
\end{subequations}
In the absence of noise, there is one fixed point $(v,\theta)=(\cos\theta_0,\theta_0)$ for each orientation $\theta_0$ of the active force. Linearizing the dynamics around any of these fixed points, one finds a zero eigenvalue for all values of $\Pi$. The destabilization threshold below which the considered fixed point is marginal and above which it is unstable is $\Pi_c(\theta_0) = \omega_1^2 / \sin^2\theta_0$. Hence all fixed points are stable for $\Pi \leq \omega_{1}^{2}$.
One recovers, as in the $N$-body problem, that the first fixed points to destabilize are those where the active force points perpendicularly to the soft direction, here in the direction $\theta_0 = \pm \pi/2$.
Conversely, in contrast with the $N$-body problem, for which it was shown that there is an upper bound for $\Pi_c$, above which all fixed points are unstable~\cite{baconnier2022selective}, here there always exist stable fixed points: more specifically the fixed points where the active force points in the soft direction, $\theta_0 = 0 \ \text{or}\ \pm\pi$, are stable for any finite value of $\Pi$.

\begin{figure}
\centering
\hspace*{-0.15cm}
 \begin{tikzpicture}
 
 \node[] at (0.0,8.4) {\includegraphics[height=4.2cm]{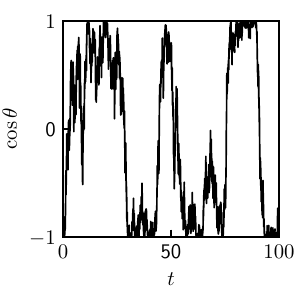}};
 \node[] at (4.2,8.4) {\includegraphics[height=4.2cm]{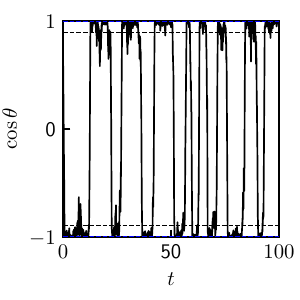}};
 
 \node[] at (-0.25,3.75) {\includegraphics[height=4.2cm]{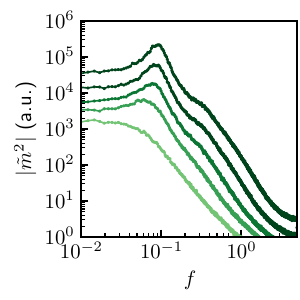}};
 \draw[->] (-0.25,3.7) -- (0.33,4.93);
 \node[] at (0.6,5.15) {\small $\Pi \nearrow$};
 
 \node[] at (4.1,3.75) {\includegraphics[height=4.2cm]{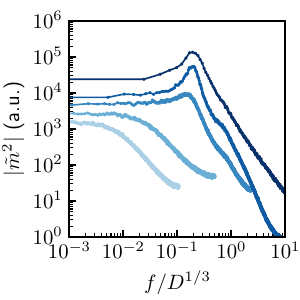}};
 \draw[->] (3.8,3.45) -- (5.2,4.85);
 \node[] at (5.55,5.05) {\small $D \nearrow$};
 
 \node[anchor=east] at (0.6,10.6) {\small (a)};
 \node[anchor=east] at (0.6,6.0) {\small (c)};
 
 \node[anchor=east] at (4.85,10.6) {\small (b)};
 \node[anchor=east] at (4.85,6.0) {\small (d)};
 
 \end{tikzpicture}
\vspace*{-0.8cm}
\caption{\small{\textbf{Dynamics of a single active particle moving along a rail:} (a, b) Temporal evolution of the polarization $m = \cos(\theta)$, below and above the instability, (a) $\Pi = 0.9 < \omega_{1}^{2}$; (b) $\Pi = 5.0 > \omega_{1}^{2}$ ($\omega_{1}^{2} = 1$, $D = 10^{-1}$). The black (resp. blue) dashed horizontal lines in (b) represent $\theta_0$ (resp. $\theta_1$). (c, d) Power spectrum of $m$ for (c) $D = 10^{-1}$ and $\Pi = 1, 2, 3, 4, 5$ and (d) $\Pi = 5$ and $D = 10^{-n}$ with $n = 4, 3, 2, 1, 0$; shifted vertically for the sake of clarity.}}
\label{fig:sp_rail_simu}
\end{figure}
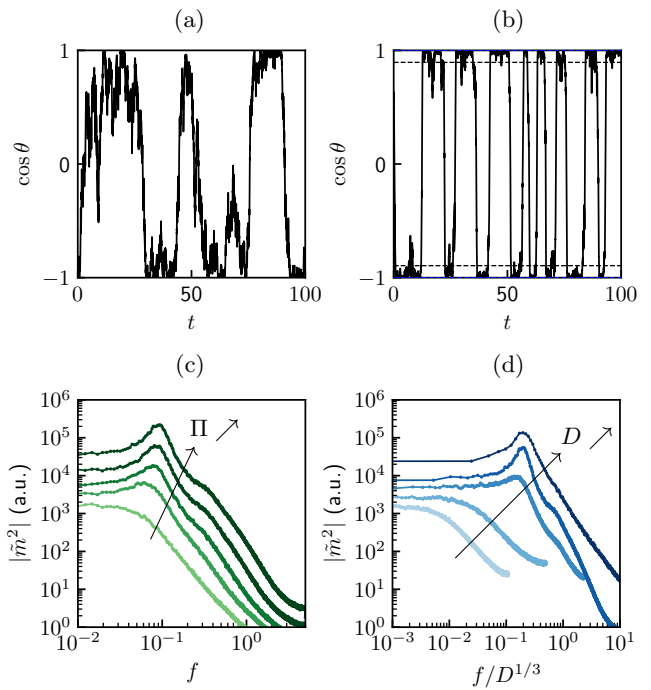

Figures~\ref{fig:sp_rail_simu}-a and ~\ref{fig:sp_rail_simu}-b display the simulated dynamics of the polarization $m=\cos(\theta)$ below and above the linear destabilization threshold of the first fixed point $\Pi_c = \omega_1^2$. Below the instability, the orientation diffuses along the continuous set of marginal fixed points.
Above the instability, the marginal fixed points lie in the ranges $[-\theta_0,\theta_0]$ and $[\pi-\theta_0,\pi+\theta_0]$, with  $\theta_0 = \arcsin(\omega_{1}/\sqrt{\Pi})$ (indicated with black dashed horizontal lines on the figure). The dynamics alternate between long periods of diffusion along one or the other subset of marginal fixed points and rapid jumps between the two subsets. An intriguing visual impression is that the stochastic jumps occur with some temporal regularity. This is confirmed when looking at the power spectrum of the polarization, $\left|\hat{m}\right|^2$, with $\hat{m}$, the Fourier transform of $m$, as shown on Figs.~\ref{fig:sp_rail_simu}-c and ~\ref{fig:sp_rail_simu}-d. When increasing $\Pi$ above the instability threshold, one observes a peak in the spectrum revealing the periodic nature of the polarization dynamics. The amplitude of the peak increases with $\Pi$, while its frequency exhibits only a weak dependence on $\Pi$. The dependence of the amplitude and location of the peak on the noise is more significant. As in the case of the $N$-particle simulations, the peak emerges for large enough noise, $D \gtrsim 10^{-2}$. The scaling of the oscillating frequency with the noise amplitude on the contrary is different from the square root dependence reported in the $N$-particle simulations. The $D^{1/3}$ scaling observed here can be explained by the dynamics at the edge of the set of marginal fixed points and is specific to the quasi-one-dimensional nature of the spatial dynamics (see Appendix~\ref{app:D13_scaling}).

\subsection{Weak noise limit}
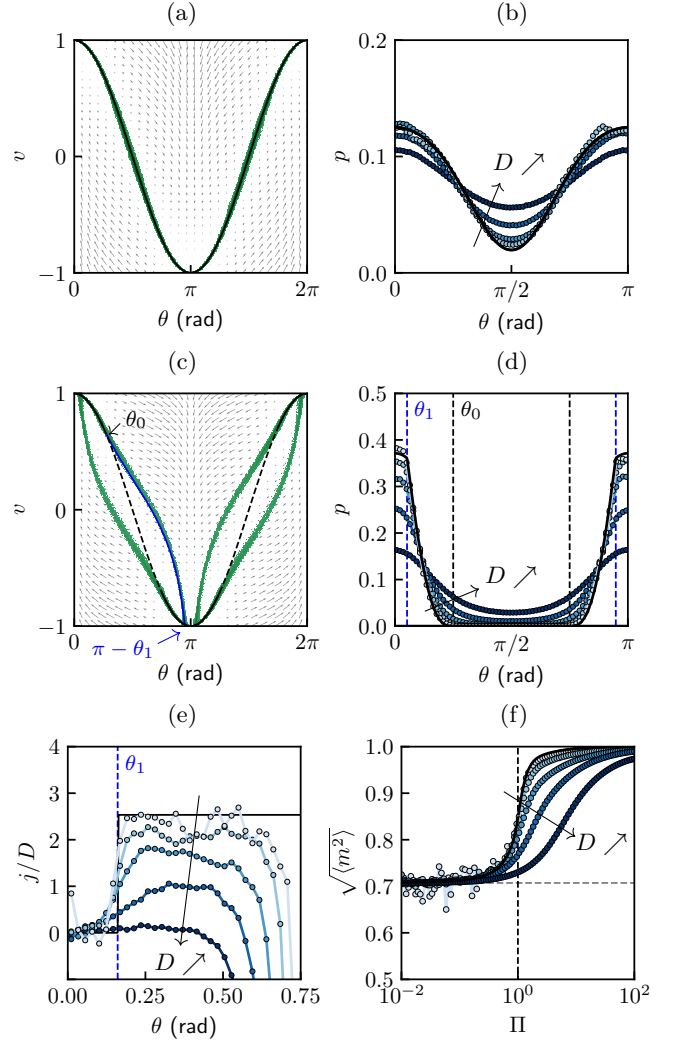
\begin{figure}
\centering
\hspace*{-0.2cm}
\begin{tikzpicture}
 
 \node[] at (0.05,8.4) {\includegraphics[height=4.3cm]{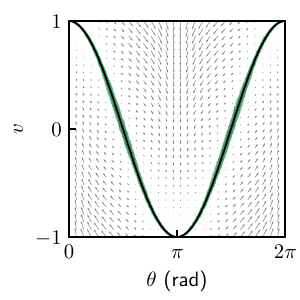}};
 \node[] at (4.4,8.4) {\includegraphics[height=4.3cm]{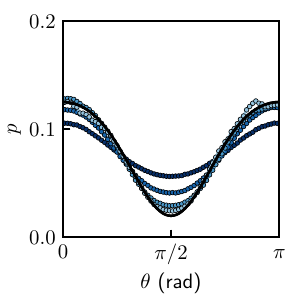}};
 
 \draw[->] (4.2,7.5) -- (4.55,8.35);
 \node[] at (4.8,8.6) {\small $D \nearrow$};
 
 \node[] at (0.05,3.7) {\includegraphics[height=4.3cm]{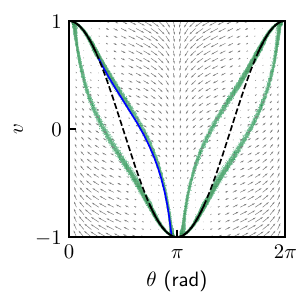}};
 
 \node[anchor=west] at (-0.55,5.2) {\footnotesize $\theta_0$};
 \draw[->] (-0.5,5.15) -- (-0.65,5.0);
 
 \node[anchor=east] at (0.08,2.15) {\footnotesize \color{blue} $\pi - \theta_1$};
 \draw[->, color=blue] (0.0,2.22) -- (0.3,2.37);
 
 \node[] at (4.4,3.7) {\includegraphics[height=4.3cm]{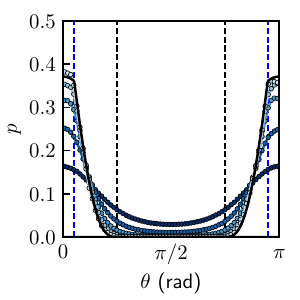}};
 
 \node[] at (3.55,5.32) {\footnotesize \color{blue} $\theta_1$};
 \node[] at (4.17,5.32) {\footnotesize $\theta_0$};
 
 \draw[->] (3.55,2.65) -- (4.27,2.95);
 \node[] at (4.7,3.10) {\small $D \nearrow$};
 
 \node[] at (0.05,-1.0) {\includegraphics[height=4.3cm]{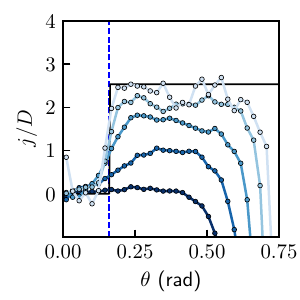}};
 
 \node[] at (-0.29,0.61) {\footnotesize \color{blue} $\theta_1$};
 \draw[->] (0.55,0.2) -- (0.3,-1.8);
 \node[] at (0.3,-2.05) {\small $D \nearrow$};
 
 \node[] at (4.4,-1.0) {\includegraphics[height=4.3cm]{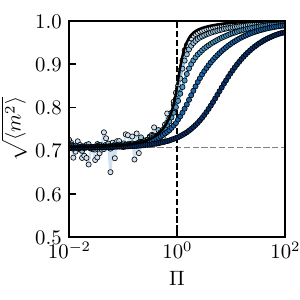}};
 
 \draw[->] (4.6,0.25) -- (5.55,-0.35);
 \node[] at (5.9,-0.45) {\small $D \nearrow$};
 
 \node[anchor=east] at (0.65,10.6) {\small (a)};
 \node[anchor=east] at (5.05,10.6) {\small (b)};
 
 \node[anchor=east] at (0.65,5.95) {\small (c)};
 \node[anchor=east] at (5.05,5.95) {\small (d)};
 
 \node[anchor=east] at (0.65,1.25) {\small (e)};
 \node[anchor=east] at (5.05,1.25) {\small (f)};
 
 \end{tikzpicture}

\vspace*{-0.4cm}
\caption{\small{\textbf{Transition to NICA for the single active particle moving along a rail in the weak noise limit:} (a, c) Distribution of the positions in phase space for $D = 10^{-4}$ and (b, d) orientations for $D \in \left[ 10^{-4}, 10^{-3}, 10^{-2}, 10^{-1}, 10^{0}\right]$, color coded from light to dark blue as the noise $D$ increases, below, $\Pi = 0.9 < \omega_{1}^{2}$ (a, b), and above, $\Pi = 2.0 > \omega_{1}^{2}$ (c, d), the instability ($\omega_{1}^{2} = 1$);
In (a-c) the black solid (resp. dashed) lines represent stable (resp. unstable) fixed points. The vector field indicates the dynamics given by Eqs. (\ref{eq:dotvth}). The blue solid line integrates the dynamics from the boundary of a component of the set of stable fixed points $\theta_0$ to the other component, with junction at $\pi - \theta_1$. 
In (b) the black solid line indicates the prediction from Eq. (\ref{eq:pred_density_theta}). In (d) the black solid line indicates the prediction from Eqs. (\ref{eq:pred_1}), (\ref{eq:pred_2}), and (\ref{eq:pred_current}); the vertical black dashed (resp. blue) line indicates $\theta_0 = \arcsin(1/\sqrt{\Pi})$ (resp. $\theta_1$, as obtained from the numerical integration of Eqs. (\ref{eq:dotvth})).
(e) Probability current in the set of stable fixed points for $\Pi = 2$ for the same values of $D$ color coded as in (b) and (d). The current is computed from Eq. (\ref{eq:current}) with the density measured in the simulations. The black solid line is the prediction from Eq. (\ref{eq:pred_current}).
(f) Root-mean-squared polarization $\sqrt{\langle m^2 \rangle}$ as a function of $\Pi$. The black solid line represents the predictions below and above the instability, respectively, Eq. (\ref{eq:pred_density_theta}) and Eqs. (\ref{eq:pred_1}), (\ref{eq:pred_2}), and (\ref{eq:pred_current}). Circles are numerical data for $D \in \left[ 10^{-4}, 10^{-3}, 10^{-2}, 10^{-1}, 10^{-0}\right]$, color coded from light to dark blue as the noise $D$ increases. The dashed horizontal line indicates the value $1/\sqrt{2}$, which corresponds to a homogeneous distribution of $\theta$.}}
\label{fig:sp_rail_weaknoise}
\end{figure}

For weak noise, the effective dynamics of the orientation can be determined analytically and provides a good intuition of the interplay between the noisy dynamics on the set of marginal fixed points and the essentially deterministic one in the rest of the phase space. 
When $\Pi \leq \omega_{1}^{2}$ and the noise is weak, the dynamics diffuse along the continuous set of marginal fixed points and a stationary distribution of the orientation, corresponding to a zero flux in phase space, can be found (Figs.~\ref{fig:sp_rail_weaknoise}-a and ~\ref{fig:sp_rail_weaknoise}-b).
When $\Pi \geq \omega_{1}^{2}$, the dynamics again diffuse along the manifold of marginal fixed points until the system reaches the boundary of one subset of fixed points, say, when $\theta$ reaches $\theta_0$ from below. It then escapes the fixed points manifold and follows a deterministic orbit that sends it to the other subset of stable fixed points, where it diffuses again until reaching the symmetric boundary and the process starts again. This induces a finite probability flux and a modified stationary distribution of the orientation (Figs.~\ref{fig:sp_rail_weaknoise}-c and ~\ref{fig:sp_rail_weaknoise}-d), which we compute below.

If $f(v,\theta,t)$ is the probability density for the variables $v$ and $\theta$, the marginal distribution of the orientation, $p(\theta,t)=\int f(v,\theta,t)\dd v$, obeys the Fokker-Planck equation (see Appendix~\ref{app:orientation})
\begin{equation}
\label{eq:fokker_planck}
\partial_t p(\theta,t) = - \partial_\theta j(\theta,t),
\end{equation}
where the current is:
\begin{multline}\label{eq:current}
j(\theta) = D \Biggl(\Biggr. \frac{\Pi \sin(\theta)\cos(\theta)}{[\omega_{1}^{2} - \Pi\sin(\theta)^2]^2}p(\theta) \\
 - \frac{\omega_1^2}{\omega_{1}^{2} - \Pi\sin(\theta)^2}\partial_\theta \left[\frac{p(\theta) }{\omega_{1}^{2} - \Pi\sin(\theta)^2}\right] \Biggl.\Biggr).
\end{multline}
When $\Pi\leq\omega_1^2$, the current is zero and the stationary density is obtained by defining $\phi(\theta)=p(\theta)/[\omega_{1}^{2} - \Pi\sin^2\theta]$, which obeys
\begin{equation}
\partial_\theta\phi = \frac{\Pi}{\omega_1^2}\sin\theta \cos\theta \phi,
\end{equation}
and integrates to $\phi=\exp(\Pi\sin^2\theta/[2\omega_1^2])$. The stationary density thus reads
\begin{equation}\label{eq:pred_density_theta}
p(\theta)\propto p^*(\theta) = [\omega_{1}^{2} - \Pi\sin^2\theta]\exp \left( \frac{\Pi}{2 \omega_1^2}\sin^2\theta \right),
\end{equation}
matching very well numerical results for small noise (Fig.~\ref{fig:sp_rail_weaknoise}-b, lightest blue curves).

When $\Pi>\omega_{1}^{2}$, the current is nonzero and proportional to $D$ as obtained from Eq.~(\ref{eq:current}). However, on the dynamical paths connecting the sets of stable fixed points, the advective force is deterministic and does not depend on $D$. In that region of phase space, the probability density itself should therefore be proportional to $D$ and decay to zero in the limit of vanishing noise.
This provides us with the appropriate boundary conditions to solve for the stationary distribution for $\theta\in[0,\theta_0]$. 
There is an incoming probability current $j$ at $\theta_1$, where the deterministic dynamics lands on the subset of stable fixed points, and by conservation of the density, the same outgoing current $j$ at $\theta_0$. The stationary probability thus satisfies Eq.~(\ref{eq:current}) with $p(\theta_0) = 0$ and $j(\theta) = DJH(\theta-\theta_1)$, where $H(x)$ is the Heaviside function.
Introducing
\begin{equation} \label{eq:pred_1}
\psi(\theta)=\frac{p(\theta)}{p^*(\theta)}=\frac{p(\theta)}{\omega_{1}^{2}-\Pi\sin^2\theta}\exp \left(-\frac{\Pi}{2 \omega_1^2}\sin^2\theta, \right),
\end{equation}
the solution can be analyzed segment by segment:
\begin{itemize}
\item For $\theta\in[0,\theta_1]$, the current is zero by symmetry and the density still follows Eq.~(\ref{eq:pred_density_theta}). Thus, $\psi(\theta)=\psi(\theta_1)$ is constant.
 \item For $\theta\in[\theta_1,\theta_0]$, $\psi(\theta)$ obeys
\begin{equation} \label{eq:pred_2}
\psi'(\theta) = -J \left[\omega_{1}^{2}-\Pi\sin^2\theta \right] \exp \left(-\frac{\Pi}{2 \omega_1^2}\sin^2\theta \right).
\end{equation}
Integrating this equation for $\theta\in[\theta,\theta_0]$, provides an explicit integral expression for $\psi(\theta)$.
\end{itemize}
Finally, using the normalization condition, $\int_0^{\theta_0}p(\theta)\dd\theta = 1/4$ we find an expression for $J$:
\begin{multline}
\label{eq:pred_current}
\frac{1}{4J} = \int_0^{\theta_1}\dd\theta p^*(\theta)\times\int_{\theta_1}^{\theta_0} \dd\theta'p^*(\theta') e^{-\Pi\sin(\theta')^2/\omega_1^2} \\
+ \int_{\theta_1}^{\theta_0}\dd\theta p^*(\theta)\int_\theta^{\theta_0} \dd\theta'p^*(\theta') e^{-\Pi\sin(\theta')^2/\omega_1^2}.
\end{multline}
The density probability for $\theta$ and the current $J$ obtained with this method compare well with the numerical simulations of the dynamics conducted with decreasing amplitude of the noise (Figs.~\ref{fig:sp_rail_weaknoise}-d and ~\ref{fig:sp_rail_weaknoise}-e).
With this explicit expression for the probability density for any activity, we compute the average squared polarization, $\sqrt{\langle m^{2} \rangle}$, as a function of $\Pi$ (Fig.~\ref{fig:sp_rail_weaknoise}-f).
We find a continuous crossover between the low- and large-activity regimes, in good agreement with numerical results in the limit of small noise. Consistently with the active ladders' simulations, the transition gets sharper as noise decreases and converges asymptotically. 

The above weak noise limit analysis captures the distributions of the orientation and the current in phase space, obtained numerically, but does not capture the emergence of a peak in the power spectrum of the magnetization. In the following, we shall see that considering the large-scale dynamics as described by coarse-grained equations, the oscillation naturally emerges through a Hopf bifurcation, as a result of the presence of noise. We shall also obtain the square root scaling of the oscillation frequency with noise reported in the simulations of the ladder. 

\section{Coarse-grained model}
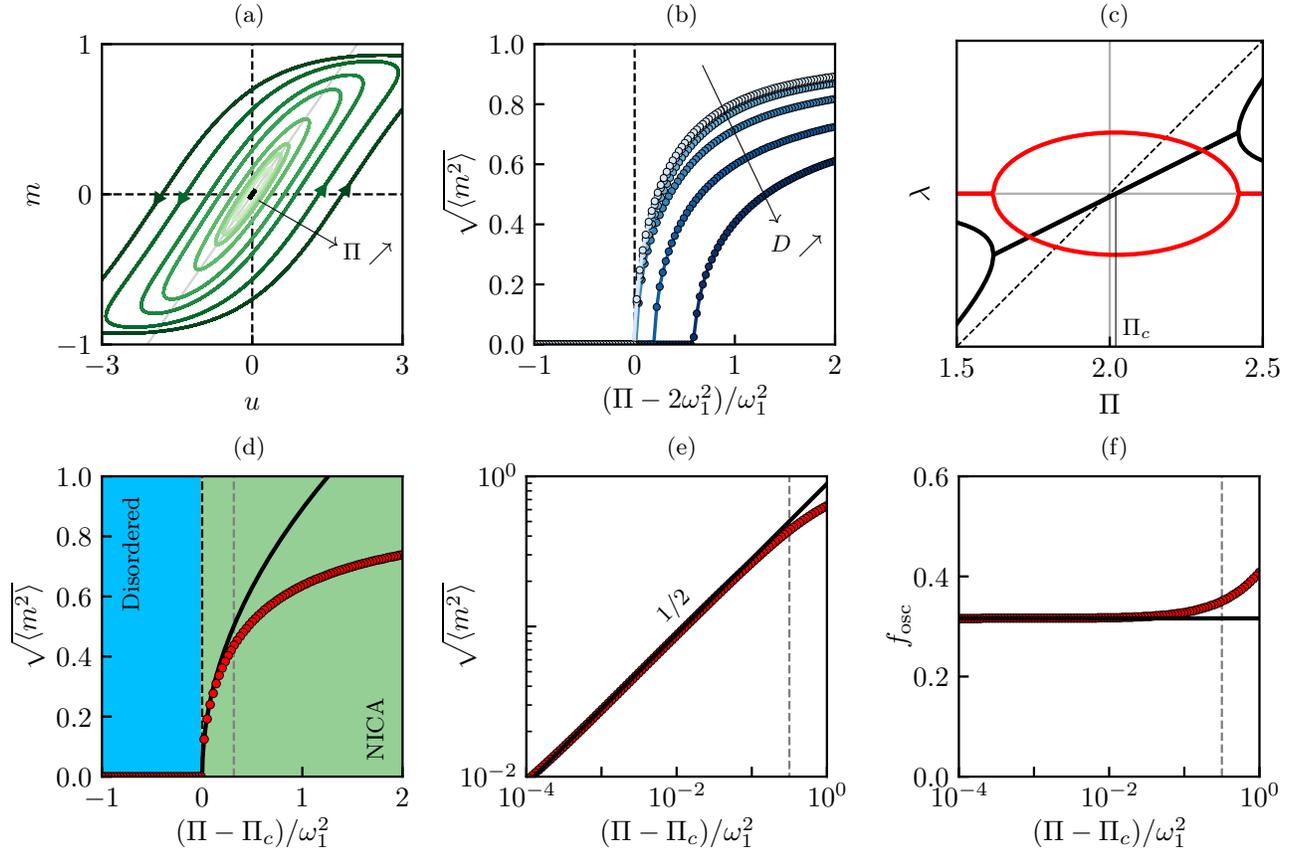
\begin{figure*}[t!]
\centering
\vspace*{0.5cm}
\begin{tikzpicture}

\node[] at (2.875,-5.75) {\includegraphics[height=5.4cm]{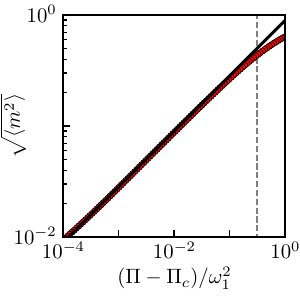}};
\node[] at (8.625,-5.75) {\includegraphics[height=5.4cm]{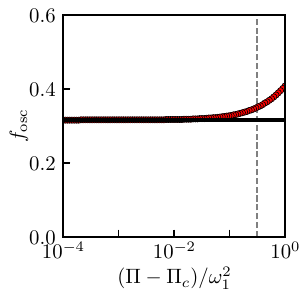}};

\node[] at (-2.875,-5.75) {\includegraphics[height=5.4cm]{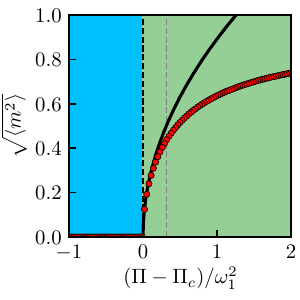}};

\node[] at (-2.875,0.0) {\includegraphics[height=5.4cm]{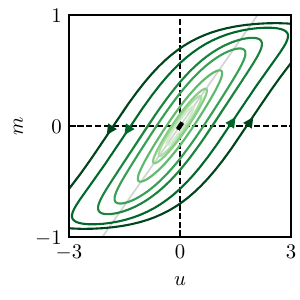}};
\node[] at (2.875,0.0) {\includegraphics[height=5.4cm]{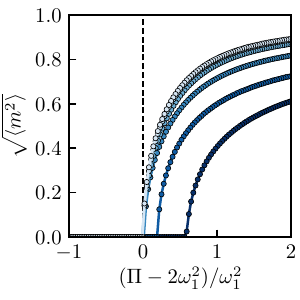}};
\node[] at (8.625,0.0) {\includegraphics[height=5.5cm]{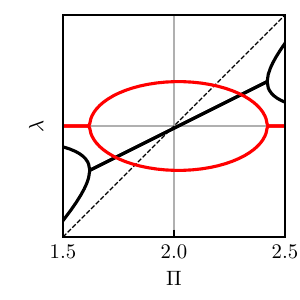}};

\draw[] (9.145,0.45) -- (9.145,-1.6);
\node[] at (9.425,-1.35) {\small $\Pi_c$};

\node[] at (-2.375,2.8) {\small (a)};
\node[] at (3.375,2.8) {\small (b)};
\node[] at (9.125,2.8) {\small (c)};

\node[] at (-2.375,-2.95) {\small (d)};

\node[] at (3.375,-2.95) {\small (e)};
\node[] at (9.125,-2.95) {\small (f)};

\draw[->] (3.650,2.15) -- (4.650,0.05);
\node[] at (4.9,-0.25) {\small $D \nearrow$};

\draw[->] (-2.25,0.35) -- (-1.2,-0.25);
\node[] at (-0.8,-0.35) {\small $\Pi \nearrow$};

\node[rotate=45] at (3.275,-5.05) {\small $1/2$};

\node[rotate=90] at (-3.95,-4.35) {\small Disordered};
\node[rotate=90] at (-0.7,-6.68) {\small NICA};

\end{tikzpicture}
\vspace*{-0.3cm}
\caption{\small{\textbf{Hopf bifurcation toward NICA at the coarse-grained level}, for fixed $\omega_{1}^{2} = 1$. 
(a) Nonlinear NICA limit cycles in the $u - m$ plane, as obtained from the numerical integration of Eqs.~(\ref{eq:toy_model_1mode}), for fixed $D = 0.1$ and $\Pi - \Pi_c \in \left[ 0.02, 0.05, 0.08, 0.2, 0.5, 0.8, 1.4, 2.0 \right] \omega_{1}^{2}$, color coded from light to dark green as $\Pi$ increases. The black marker at the origin represents the disordered fixed point, and the gray solid line corresponds to the vector $(a,1)$, as given by Eq. (\ref{eq:multiple_scale_zero}).
(b) Amplitude of oscillations along $m$ as a function of the normalized distance to threshold $(\Pi - 2\omega_{1}^{2})/\omega_{1}^{2}$ for different noise amplitude $D \in \left[ 3.10^{-1}, 10^{-1}, 10^{-2}, 10^{-3}, 10^{-4}, 10^{-5} \right]$, color coded from light to dark blue as $D$ increases. Markers are obtained from the numerical integration of Eqs.~(\ref{eq:toy_model_1mode}).
(c) Linear stability analysis of the disordered phase in the coarse-grained model: solutions of  Eq. (\ref{eq:eigenvalue_problem}) as a function of $\Pi$, for $D = 10^{-2}$. Black (resp. red) curves represent the real (resp. imaginary) parts of the solutions. The black dashed line represents the nonzero solution for $D = 0$, i.e., $\lambda = \Pi/2 - \omega_{1}^{2}$.
(d) Amplitude of oscillations along $m$ as a function of the normalized distance to threshold $(\Pi - \Pi_c)/\omega_{1}^{2}$, for $D = 0.1$. The solid black line represents the prediction from Eq. (\ref{eq:AmplitudeEquationR_cg1mode_main}), and the red markers represent the numerical results, as obtained from the numerical integration of Eqs. (\ref{eq:toy_model_1mode}). The dashed gray line highlights the timescale separation's validity threshold in activity. 
(e) Log-log representation of (d) close to the threshold, with the same conventions. 
(f) Frequency of the NICA limit cycle as a function of the normalized distance to threshold $(\Pi - \Pi_c)/\omega_{1}^{2}$, as obtained from the largest peak of the $m(t)$ Fourier transform. Same conventions as (d) and (e).}}
\label{fig:CG_NICA}
\end{figure*}

At large scale, one introduces the spatially continuous fields $\uu(\rr, t)$ and $\mm(\rr , t)$, which respectively describe the local averages of the microscopic displacements $\boldsymbol{u}_i$ and polarizations $\hat\nn_i$.
Their dynamics are described by the coarse-grained equations obtained in~\cite{baconnier2022selective}: 
\begin{subequations} \label{eq:coarse_grained_relax}
\begin{align}
 \partial_{t} \boldsymbol{u} &= \Pi \boldsymbol{m} + \boldsymbol{F}^\mathrm{el}\left[ \boldsymbol{u} \right], \label{eq1:coarse_grained_relax} \\
 \partial_{t}\boldsymbol{m} &= ( \boldsymbol{m} \times \boldsymbol{F}^\mathrm{el} ) \times \boldsymbol{m} + \frac{1 - \boldsymbol{m}^{2}}{2} \partial_{t} \boldsymbol{u} - D \boldsymbol{m}, \label{eq2:coarse_grained_relax}
\end{align}
\end{subequations}
where the relaxation term $-D \boldsymbol{m}$ stems from the coarse-graining of the angular noise, and where the elastic force $\boldsymbol{F}^\mathrm{el}\left[ \boldsymbol{u}  \right] = \text{div} \sigma$ is given by the standard constitutive relation for continuum linear elasticity, i.e., Hooke's law, $\sigma = \frac{E}{1 + \nu} \left( \varepsilon + \frac{\nu}{1 - 2\nu} \Tr(\varepsilon) \mathbb{I} \right)$, where $\sigma$ and $\varepsilon$ are the stress and strain tensors, respectively, and $E$ and $\nu$ are the Young modulus and the Poisson ratio of the elastic material, respectively.
Note that such an elastic sheet has an infinity of normal modes, and we keep denoting $\omega_1^2$ the energy of the lowest-energy one.

In the absence of noise, $D=0$, any active force field balancing the elastic forces, $\Pi \mm(\rr, t) = - \boldsymbol{F}^\mathrm{el}\left[ \boldsymbol{u} \right]$, is a stationary solution of the coarse-grained equations. 
This is not the case in the presence of noise, however small is $D$. The noise enforces the existence of a single fixed point ($\uu = 0$, $\mm = 0$), which corresponds to the homogeneous disordered state.  
Large enough elasto-active coupling $\Pi$ leads the destabilization of this fixed point into a limit cycle (see the numerical integration of the projected equations defined below in Fig.~\ref{fig:CG_NICA}-a), the amplitude of which seems to grow continuously from zero at onset (Fig.~\ref{fig:CG_NICA}-b). 
The linear stability analysis of the fixed point  ($\uu = 0$, $\mm = 0$) leads to the following eigenvalue problem (see Supplementary Information of \cite{baconnier2022selective}, Sec. 6.5):
\begin{equation} \label{eq:eigenvalue_problem}
	\lambda^{2} - \lambda \left( \frac{\Pi}{2} - \omega_{1}^{2} - D \right) + D \omega_{1}^{2} = 0.
\end{equation}
The real part of the two eigenvalues increases from negative values at small $\Pi$, where the disordered phase is stable, and becomes positive for $\Pi > \Pi_c = 2\left( \omega_1^2 + D \right)$. For any finite $D$, the two eigenvalues collide on both sides of the instability threshold, leading to complex conjugate values at the instability threshold (Fig.~\ref{fig:CG_NICA}-c), which is the hallmark of a Hopf bifurcation. From Eq.~(\ref{eq:eigenvalue_problem}), one easily finds that the frequency at onset scales like $\sqrt{D}$. 
Finally, the supercritical nature of the transition, anticipated from (Fig.~\ref{fig:CG_NICA}-b) is confirmed by the normal form of the bifurcation, obtained from a multiple scale analysis~\cite{jakobsen2013introduction}, which we summarize below (see Appendix~\ref{app:multiple_scale_analysis} for the complete derivation).

Starting with the projection of the coarse-grained equations along the lowest-energy mode,  
\begin{subequations} \label{eq:toy_model_1mode}
\begin{align}
 \partial_{t} u &= \Pi m  - \omega_{1}^{2} u, \label{eq1:toy_model_1mode} \\
 \partial_{t} m &= \frac{1 - m^{2}}{2} \left( \Pi m - \omega_{1}^{2} u \right) - D m. \label{eq2:toy_model_1mode}
\end{align}
\end{subequations}
one uses the fact that, close to the threshold, the dynamics along the unstable manifold is slow as compared to that in all other directions in phase space. Introducing the small parameter $\epsilon = (\Pi-\Pi_c)/\Pi_c$, one defines the slow timescale $T = \epsilon t$ and prescribes the scalings $u \rightarrow \sqrt{\epsilon}\,u(t,T)$ and $m \rightarrow \sqrt{\epsilon}\,m(t,T)$, in light of the parity symmetry of the problem.  One then looks for solutions of the form:
\begin{subequations} \label{eq:mutiscaling_cg}
\begin{align}
 u(t,T)  &= u_0(t,T) + \epsilon u_1(t,T) + \hdots \label{eq1:mutiscaling_cg} \\
 m(t,T)  &= m_0(t,T) + \epsilon m_1(t,T) + \hdots \label{eq2:mutiscaling_cg}
\end{align}
\end{subequations}
solving order by order in $\epsilon$. At zeroth order, the solutions satisfy a system of linearly coupled ODEs and read:
\begin{equation} \label{eq:multiple_scale_zero}
 \begin{pmatrix}
  u_0(t,T) \\
  m_0(t,T)
 \end{pmatrix}
= A(T) \begin{pmatrix}
  a \\
  1
 \end{pmatrix} e^{i\Omega t} +
A^{\star}(T) \begin{pmatrix}
  a^{\star} \\
  1
 \end{pmatrix} e^{-i\Omega t},
\end{equation}
where $a = 2\left( 1 - i \sqrt{\mu} \right)$, $\mu = D/\omega_{1}^{2}$, and $\Omega = \omega_{1} \sqrt{D}$. Further solving at first order and imposing the so-called Fredholm alternative as a solvability condition, one finally obtains the amplitude equation for the complex amplitude $A$ and its modulus $R=\left|A\right|$:
\begin{equation} \label{eq:AmplitudeEquationR_cg1mode_main}
\frac{dR}{dT} = \epsilon \frac{2 + \mu}{4} R - 3 D \frac{1+\mu}{2} R^3.
\end{equation}
This normal form describes a supercritical Hopf bifurcation, for which the solution $R=0$, linearly stable for $\epsilon<0$, destabilizes into a limit cycle of amplitude $R\sim\sqrt{\epsilon}$ and finite frequency $\Omega\sim\sqrt{D}$ for $\epsilon>0$.
The range of validity of the analysis is provided by the separation of the slow $\tau_{\rm relax} \propto \epsilon^{-1}$ and fast $\tau_{\rm osc} = \Omega^{-1}$ time scales, that is $\epsilon \ll \omega_1\sqrt{D}$. 
We note that the coefficient of the third-order nonlinear saturating term, the oscillating frequency, and the range of validity vanish in the limit of zero noise, pointing at the singularity of this limit.

Figures~\ref{fig:CG_NICA}-d, ~\ref{fig:CG_NICA}-e, and ~\ref{fig:CG_NICA}-f provide a comparison of the prediction from the multiple-scale analysis, Eq.~(\ref{eq:AmplitudeEquationR_cg1mode_main}), to the numerical integration of the coarse-grained equations~(\ref{eq:toy_model_1mode}). The amplitude of oscillation found numerically matches very well the prediction close to the threshold $\Pi = \Pi_c$, with a clear square root power law (Fig.~\ref{fig:CG_NICA}-e). The oscillation frequency $f_\textrm{osc}$ of the NICA limit cycle, as given by the largest peak of the Fourier transform of $m(t)$, is in agreement with a Hopf bifurcation at $\Omega = \sqrt{D\omega_{1}^{2}}$ (Fig.~\ref{fig:CG_NICA}-f).

The results obtained from the coarse-grained model and multiple scale analysis are also in good qualitative agreement with the transition found numerically for the $N$-particle active ladders. For not too large noise, the thresholds $\Pi_c = 2(\omega_{0}^{2} + D)$ indicated with the vertical shaded blue lines in Fig.~\ref{fig:simulation_hopf}-c are good estimates of the inflection points marking the crossover in the transversal polarity from the disordered state to the oscillating regime.
The scaling of the oscillation frequency in $\sqrt{D}$ at onset, reported in Figs.~\ref{fig:simulation_hopf}-d is correctly predicted.  Farther from the threshold, we finally note a gentle increase of the oscillation frequency computed from the coarse-grained equations (Fig.~\ref{fig:CG_NICA}-f), also in agreement with the numerical simulations of the $N$-particle active ladders (Figs.~\ref{fig:simulation_hopf}-d,f).
\section{Discussion}
The noise-induced collective actuation is a rich, yet simple enough, example of the role of noise in a nonequilibrium system.
The phase space structure and the noise combine to promote regular oscillations between two regions of the phase space, which contrasts with the equilibrium picture of stochastic jumps between two wells of a free energy landscape. 

This picture is well captured by the model of a single particle moving along a rail. In this case, the phase space structure is exactly known with a clear separation of the set of marginal fixed points in two disconnected subsets connected by deterministic trajectories, akin to heteroclinic orbits, in the language of dynamical systems. The diffusion within one subset of marginally stable fixed points is thus followed by a quasi-deterministic run along these orbits towards the other subset.  The system does not have to overcome any barrier to do so, and the dynamics is therefore very different from an activated one. This simple model captures the spontaneously broken symmetry of the low-activity \textit{frozen-polarized} phase, while properly considering the stochastic nature of angular noise in the polarity dynamics and polarization reversals. However, the activity threshold predicted, $\Pi_c = \omega_{1}^{2}$, conspicuously deviates from that reported in the $N$-particle simulations by a factor $2$, and the scaling of the oscillation frequency with noise is in $D^{1/3}$ instead of the reported $D^{1/2}$. Also, the system is doomed to exhibit a finite polarization in the nonoscillating phase, while the latter averages to zero in the disordered phase of the $N$-particle system.

In contrast, the coarse-grained model with finite noise offers the possibility of a disordered phase with vanishing polarization. In this case, we explicitly find a supercritical Hopf bifurcation between a fully disordered, frozen phase and a nonlinear NICA limit cycle at $\Pi_c = 2 (\omega_{1}^2 + D)$, with a frequency at the onset of oscillations scaling like $\sqrt{D}$, both matching well with numerical simulations of active ladders. Remarkably, the derivation of the NICA limit cycle does not require specifying the lowest-energy mode's geometry, emphasizing the generality of NICA. Therefore, in the presence of noise, we expect that large enough activity leads to the emergence of spontaneous oscillations along the lowest-energy mode in any active elastic structures whose normal mode spectrum is gapped enough.

Collective actuation in elastic structures was first studied in the absence of noise~\cite{baconnier2022selective, baconnier2023discontinuous}. In that case, the oscillations take place on a pair of modes that are nonlinearly selected. The dynamics projected on this pair of modes is equivalent to that of a single self-aligning polar particle in a two-dimensional harmonic potential, the stiffnesses of which are given by the energies of the selected modes. In this context, noise-induced collective actuation appears to be the limiting case of collective actuation, when the two selected modes are gapped, and the elasto-active feedback alone cannot give rise to oscillations. We thus infer that, in the presence of noise, when the two selected modes are nondegenerated, the noiseless transition is likely to be preceded by a noise-induced collective actuation regime with condensation on the one of the two selected modes with the lowest energy. This could explain the spontaneous oscillations of an active solid along its lowest-energy mode as reported in~\cite{henkes2011active} in the context of jammed active solids or in Vertex models~\cite{barton2017active, petrolli2019confinement}. More generally, a better understanding of the crossover between these two regimes of collective actuation would certainly be of interest.

Let us comment on the connections between collective actuation in active solids and synchronization of coupled oscillators~\cite{kuramoto1975self, mirollo1990synchronization, strogatz1993coupled, pikovsky2001universal, strogatz2004sync,
acebron2005kuramoto, o2017oscillators}. Indeed, such systems also exhibit a disordered phase for small enough coupling, and an increasing fraction of synchronized oscillators for couplings
larger than a critical value. There are, however, two important differences to be made with collective actuation. First, the active units are not oscillators in that they don't have an intrinsic frequency; they push and re-orient upon being displaced,  but they do not rotate spontaneously. The oscillations and the synchronization emerge concomitantly. Second, in the specific case of NICA, it is the noise and not the coupling strength that promotes the synchronized oscillations.

Finally, a transition to spontaneous oscillations via a Hopf bifurcation was observed in another system of elastically-connected self-propelled robots~\cite{zheng2023self}, but in that case the instability results from the coupling between activity and elastic nonlinearities. Extending the current framework where elasticity is purely linear to the nonlinear regime is a promising route to decipher the mechanisms at the origin of spontaneous oscillations observed in dense living systems~\cite{henkes2011active, serra2012mechanical, deforet2014emergence, barton2017active, petrolli2019confinement, peyret2019sustained, liu2021viscoelastic, xu2023autonomous}. First, in this context, noise can trigger plastic spatial rearrangements of the structure, affecting its vibrational properties~\cite{henkes2012extracting}. Second, investigating the influence of quenched disorder on the transition to collective actuation~\cite{dotsenko1995critical} could provide valuable insights into the interplay between system and environmental variability~\cite{bechinger2016active}, and the emergence of collective oscillations.

\begin{acknowledgments}
We thank A. C. Maggs for spotting the $D^{1/3}$ scaling in the power spectrum of the single particle and providing us with an argument to explain it. We acknowledge financial support from Ecole Doctorale ED564 Physique en Ile de France for P.B.'s Ph.D. grant.
\end{acknowledgments}

\appendix

\begin{widetext}

\section{Argument for the $D^{1/3}$ scaling of the power spectrum of a single particle}
\label{app:D13_scaling}

In this appendix, we present an argument for the $D^{1/3}$ scaling arising in the power spectrum of a single particle (Fig.~\ref{fig:sp_rail_simu}).
The fixed point at the edge of the marginal fixed points is close to its instability threshold. The dynamics around this point is therefore extremely slow. We assume that the particle spends most of the time ``turning around'' the edge of the marginal fixed points, and expand the dynamics around this point ($\theta=\theta_0$, $v=\cos(\theta_0)$).

Writing $\theta=\theta_0+\phi$, $v=\cos(\theta_0)+x$ with $\phi, x \ll 1$ in Eq.~(\ref{eq:dotvth}), we obtain
\begin{align}
\dot x&=-\sin(\theta_0)\phi-x,\\
\dot\phi & = \frac{\Pi}{\omega_1^2}\sin(\theta_0)[\sin(\theta_0)\phi+x]+\sqrt{\frac{2D}{\omega_1^2}}\xi.
\end{align}
We introduce the variable $y=\sin(\theta_0)\phi+x$, which follows
\begin{equation}
\dot y=\left[\frac{\Pi}{\omega_1^2}\sin(\theta_0)^2-1\right] y + \sin(\theta_0)\sqrt{\frac{2D}{\omega_1^2}}\xi = \sqrt{\frac{2D}{\Pi}}\xi,
\end{equation}
where we have used $\sin(\theta_0)=\omega_1/\sqrt{\Pi}$.
The variable $y$ thus follows purely Brownian dynamics.

Assuming that the system starts at the edge of the stable fixed points at $t=0$, then, 
\begin{equation}
\langle y(t)^2 \rangle = \frac{2D t}{\Pi}.
\end{equation}
We now focus on $x$, which follows $\dot x=-y$, and compute its variance:
\begin{equation}
\langle x(t)^2 \rangle = \frac{2D t^3}{3\Pi}.
\end{equation}

If a distance $\ell$ from the point $(\cos(\theta_0), \theta_0)$ has to be reached, after which the fast deterministic dynamics takes over, it takes a time $\tau = \ell^{2/3} (\Pi/D)^{1/3}$, leading to a characteristic frequency
\begin{equation}
f\sim D^{1/3}.
\end{equation}    

\section{Equation for the orientation for the single particle}
\label{app:orientation}

To obtain the effective evolution equation of the marginal distribution of the orientation $p(\theta,t)$ in the limit of weak noise, we start with the Fokker-Planck equation for the phase space density $f(v, \theta,t)$ and use a technique reminiscent of the adiabatic elimination of a fast variable (\cite{Risken1996}, Sec. 8.3). 
Here, the fast variable is the deviation from the manifold of marginally stable fixed points.

We start with the Fokker-Planck equation for the phase-space density $f(v, \theta,t)$ associated to with Langevin equations (\ref{eq:dotv}) and (\ref{eq:dotth_v}):
\begin{equation}\label{eq:fp_vth}
\partial_t f(v,\theta,t) = -\partial_v j_v(v,\theta,t)-\partial_\theta j_\theta(v,\theta,t),
\end{equation}
where the currents are
\begin{align}
j_v(v,\theta) & = [\cos(\theta)-v]f(v,\theta),\\
j_\theta(v,\theta) & = \tilde\Pi\sin(\theta)[v-\cos(\theta)]f(v,\theta)-\tilde D\partial_\theta f(v,\theta).\label{eq:fp_vt_curt}
\end{align}
We have defined $\tilde \Pi=\Pi/\omega_1^2$ and $\tilde D=D/\omega_1^2$; we omit the tilde in the following.

We are interested in the equation satisfied by the marginal density $p(\theta)=\int f(v,\theta)\dd v$. 
Integrating Eq.~(\ref{eq:fp_vth}) over $v$, we obtain its evolution equation
\begin{equation}
\partial_t p(\theta,t) = -\partial_\theta \bar j_\theta(\theta, t),
\end{equation}
where we have defined the integrated current $\bar j_\theta(\theta) = \int j_\theta(v, \theta)\dd v$.
The integrated current can be obtained from Eq.~(\ref{eq:fp_vt_curt}), 
\begin{equation}\label{eq:fp_th_current_moments}
\bar j_\theta(\theta) = \Pi\sin\theta f_1(\theta)-D\partial_\theta f_0(\theta),
\end{equation}
where we have defined the moments
\begin{equation}
f_n(\theta) = \int[v-\cos(\theta)]^nf(v,\theta)\dd v,
\end{equation}
for $n\geq 0$; note that $p(\theta)=f_0(\theta)$.

To compute the moments, we assume that the phase-space density is stationary:
\begin{equation}\label{eq:fp_vt}
-\partial_v \left([v-\cos(\theta)]f(v,\theta) \right) =
-\partial_\theta \left( \Pi\sin(\theta)[v-\cos(\theta)]f(v,\theta)\right)+ D\partial_\theta^2 f(v,\theta).
\end{equation}
Multiplying Eq.~(\ref{eq:fp_vt}) by $[v-\cos(\theta)]^n$ and integrating, we get
\begin{equation}\label{eq:fnth_0}
n f_n(\theta) = 
-\Pi\int [v-\cos(\theta)]^n \partial_\theta \left(\sin(\theta)[v-\cos(\theta)]f(v,\theta)\right)\dd v+D\int [v-\cos(\theta)]^n\partial_\theta^2 f(v,\theta)\dd v,
\end{equation}
where we have integrated by parts to get the left-hand side, assuming that the boundary terms vanish.
We now use that, for a generic function $g(\theta)$,
\begin{equation}
[v-\cos(\theta)]^n \partial_\theta g(\theta) = \partial_\theta \left([v-\cos(\theta)]^n g(\theta) \right) - n\sin(\theta) [v-\cos(\theta)]^{n-1} g(\theta),
\end{equation}
to rewrite the second term in the r.h.s. of Eq.~(\ref{eq:fnth_0}):
\begin{equation}
\int [v-\cos(\theta)]^n \partial_\theta \left(\sin(\theta)[v-\cos(\theta)]f(v,\theta)\right)\dd v
= \partial_\theta \left[\sin(\theta) f_{n+1}(\theta) \right]-n\sin(\theta)^2 f_n(\theta).
\end{equation}
For the third term in the  r.h.s. of Eq.~(\ref{eq:fnth_0}), we need
\begin{align}
[v-\cos(\theta)]^n\partial_\theta^2 f(v,\theta) & = \partial_\theta^2 \left([v-\cos(\theta)]^n f(v,\theta) \right) - 2\partial_\theta \left( [v-\cos(\theta)]^n\right)\partial_\theta f(v,\theta) - \partial_\theta^2 \left( [v-\cos(\theta)]^n\right)f(v,\theta)\\
& = \partial_\theta^2 \left([v-\cos(\theta)]^n f(v,\theta) \right) - 2\partial_\theta \left(\partial_\theta \left( [v-\cos(\theta)]^n\right)f(v,\theta) \right) + \partial_\theta^2 \left( [v-\cos(\theta)]^n\right)f(v,\theta)\\
& = \partial_\theta^2 \left([v-\cos(\theta)]^n f(v,\theta) \right) - 2n\partial_\theta \left(\sin(\theta)  [v-\cos(\theta)]^{n-1}f(v,\theta) \right)\nonumber\\
& \qquad  + \left(n\cos(\theta)[v-\cos(\theta)]^{n-1}+ n(n-1)\sin(\theta)^2[v-\cos(\theta)]^{n-2}\right)f(v,\theta)
\end{align}
Integrating over $v$ thus gives
\begin{equation}
\int [v-\cos(\theta)]^n\partial_\theta^2 f(v,\theta)\dd v = 
\partial_\theta^2 f_n(\theta)-2n\partial_\theta \left[\sin(\theta)f_{n-1}(\theta) \right]+n\cos(\theta)f_{n-1}(\theta)+n(n-1)\sin(\theta)^2f_{n-2}(\theta).
\end{equation}
We can now write all the terms of Eq.~(\ref{eq:fnth_0}) with the moments $f_n(\theta)$:
\begin{multline}\label{eq:fnth}
n f_n(\theta) = -\Pi  \partial_\theta \left[\sin(\theta) f_{n+1}(\theta) \right]+n\Pi\sin(\theta)^2 f_n(\theta)\\ +D \left[ \partial_\theta^2 f_n(\theta)-2n\partial_\theta \left[\sin(\theta)f_{n-1}(\theta) \right]+n\cos(\theta)f_{n-1}(\theta)+n(n-1)\sin(\theta)^2f_{n-2}(\theta)\right];
\end{multline}
we obtain a hierarchy of equations.

For $n=0$, $n=1$ and $n=2$, Eq.~(\ref{eq:fnth}) reduces to:
\begin{align}
0&=-\Pi\partial_\theta[\sin(\theta)f_1(\theta)]+D\partial_\theta^2 f_0(\theta),\label{eq:f0th}\\
[1-\Pi\sin(\theta)^2]f_1(\theta) &= -\Pi\partial_\theta[\sin(\theta)f_2(\theta)]+D \left[\partial_\theta^2f_1(\theta)-2\partial_\theta \left[\sin(\theta)f_{0}(\theta) \right]+\cos(\theta)f_{0}(\theta) \right],\label{eq:f1th}\\
2[1-\Pi\sin(\theta)^2]f_2(\theta)&=-\Pi  \partial_\theta \left[\sin(\theta) f_{3}(\theta) \right]+D \left[ \partial_\theta^2 f_2(\theta)-4\partial_\theta \left[\sin(\theta)f_{1}(\theta) \right]+2\cos(\theta)f_{1}(\theta)+2\sin(\theta)^2f_{0}(\theta)\right].\label{eq:f2th}
\end{align}
To close the hierarchy, we consider the situation where $D\to 0$.
The relations above are compatible with the scalings $f_0\sim D^0$, $f_1\sim f_2\sim D$, $f_{n>2}=o(D)$.
Restricting ourselves to the leading orders, the relations (\ref{eq:f1th}) and (\ref{eq:f2th}) read
\begin{align}
\left[1-\Pi\sin(\theta)^2\right]f_1(\theta) &= -\Pi\partial_\theta[\sin(\theta)f_2(\theta)]+D \left[-2\partial_\theta \left[\sin(\theta)f_{0}(\theta) \right]+\cos(\theta)f_{0}(\theta) \right],\label{eq:f1th_D}\\
\left[1-\Pi\sin(\theta)^2\right]f_2(\theta)&=D \sin(\theta)^2f_{0}(\theta).\label{eq:f2th_D}
\end{align}
From these equations we deduce $f_2(\theta)$ and then
\begin{equation}
f_1(\theta) = -\frac{D}{1-\Pi\sin(\theta)^2} \left[\partial_\theta\left(\frac{\sin(\theta)}{1-\Pi\sin(\theta)^2}f_0(\theta)\right)+ \sin(\theta)\partial_\theta f_0(\theta)\right].
\end{equation}
Finally, the integrated current (\ref{eq:fp_th_current_moments}) is
\begin{equation}
\bar j(\theta) = D \left[\frac{\Pi\sin(\theta)\cos(\theta)}{[1-\Pi\sin(\theta)^2]^2}p(\theta)-\frac{1}{1-\Pi\sin(\theta)^2}\partial_\theta \left(\frac{p(\theta)}{1-\Pi\sin(\theta)^2} \right) \right].
\end{equation}

Inserting back the expressions for $\tilde D$ and $\tilde \Pi$, we get
\begin{equation}
\bar j(\theta) = D \left[\frac{\Pi\sin(\theta)\cos(\theta)}{[\omega_1^2-\Pi\sin(\theta)^2]^2}p(\theta)-\frac{\omega_1^2}{\omega_1^2-\Pi\sin(\theta)^2}\partial_\theta \left(\frac{p(\theta)}{\omega_1^2-\Pi\sin(\theta)^2} \right) \right].
\end{equation}

\section{Multiple scale analysis} 
\label{app:multiple_scale_analysis}
We start from Eqs.~(\ref{eq:toy_model_1mode}). Then, we provide the derivation of the nonlinear NICA limit cycle's explicit expression, as the elasto-active coupling $\Pi$ exceeds the disordered phase stability threshold, $\Pi_c = 2(\omega_{1}^{2} + D)$. \\
\paragraph{Scaling variables.} We consider the elasto-active coupling very close to the disordered phase's stability threshold: $\Pi = \Pi_c + \delta$, where $\delta = \lambda \Delta$, with $\lambda$ a small parameter; and introduce a slow timescale $T = \lambda t$. We propose the scalings $U = \sqrt{\lambda} U(t,T)$ and $m = \sqrt{\lambda} m(t,T)$, and look for solutions of the form:
\begin{subequations} \label{eq:mutiscaling_cg}
\begin{align}
 U(t,T)  &= U_0(t,T) + \lambda U_1(t,T) + \hdots \label{eq1:mutiscaling_cg} \\
 m(t,T)  &= m_0(t,T) + \lambda m_1(t,T) + \hdots \label{eq2:mutiscaling_cg}
\end{align}
\end{subequations}
\paragraph{Perturbation.} Re-injecting Eqs. (\ref{eq:mutiscaling_cg}) into Eqs. (\ref{eq:toy_model_1mode}), we next separate the different orders in $\lambda$:
\begin{itemize}

 \item At zeroth order in $\lambda$, we find:
\begin{equation} \label{eq:mutiscaling_cg_0th}
 \frac{\partial}{\partial t} \begin{pmatrix}
  U_0(t,T) \\
  m_0(t,T)
 \end{pmatrix}
= \begin{pmatrix}
  -\omega_{1}^{2} & \Pi_c \\
  - \omega_{1}^{2}/2 & \frac{\Pi_{c}}{2} - D
 \end{pmatrix}
\begin{pmatrix}
  U_0(t,T) \\
  m_0(t,T)
 \end{pmatrix} = \mathbb{D} \begin{pmatrix}
  U_0(t,T) \\
  m_0(t,T)
 \end{pmatrix}.
\end{equation}
The eigenvalues of $\mathbb{D}$ are $\pm i \Omega$, where $\Omega = \sqrt{D \omega_{1}^{2}}$. Imposing real solutions, we find:
\begin{equation}
 \begin{pmatrix}
  U_0(t,T) \\
  m_0(t,T)
 \end{pmatrix}
= A(T) \begin{pmatrix}
  a \\
  1
 \end{pmatrix} e^{i\Omega t} +
A^{\star}(T) \begin{pmatrix}
  a^{\star} \\
  1
 \end{pmatrix} e^{-i\Omega t},
\end{equation}
where $a = 2\left( 1 - i \frac{\sqrt{D}}{\omega_0} \right)$ ; where the complex number $A(T)$ depends on the slow timescale $T$, and where the two vectors $^{t}(a,1)$ and $^{t}(a^{\star},1)$ are, respectively, the eigenvectors associated with the eigenvalues $i\Omega$ and $-i\Omega$.
 \item At first order in $\lambda$, we find
 \begin{equation} \label{eq:mutiscaling_cg_1st}
 \frac{\partial}{\partial t} \begin{pmatrix}
  U_1(t,T) \\
  m_1(t,T)
 \end{pmatrix}
 = \mathbb{D} \begin{pmatrix}
  U_1(t,T) \\
  m_1(t,T)
 \end{pmatrix}
 + \begin{pmatrix}
\Delta m_0 - \frac{\partial U_0}{\partial T} \\
\frac{1}{2} \Delta m_0 - \frac{\Pi_c}{2}m_0^3 + \frac{\omega_{0}^2}{2}m_0^2 U_0 - \frac{\partial m_0}{\partial T}
\end{pmatrix},
\end{equation}
where the matrix $\mathbb{D}$ is the same as in Eqs. (\ref{eq:mutiscaling_cg_0th}). There is no need to solve explicitly for $U_1$ and $m_1$: some terms drive the system at the resonance frequency $\omega = \pm \Omega$, and the system will generally have no solution. It will only have a solution, leading to a bounded solution for Eq. (\ref{eq:mutiscaling_cg_1st}), if the right-hand side satisfies a certain constraint. This constraint we get from the Fredholm alternative theorem.
\end{itemize}
\paragraph{Fredholm alternative theorem.} The resonant terms of the right-hand side of Eqs. (\ref{eq:mutiscaling_cg_1st}) must be orthogonal to any vector of the kernel of the matrix $\left(i\Omega \mathbb{I} - \mathbb{D} \right)^{\star}$. A basis for this subspace is the vector ${}^{t}(\frac{1}{2}\frac{\omega_{1}^{2} + i\omega_{0}\sqrt{D}}{\omega_{1}^{2} + D},1)$. Thus, the solvability condition reads:
\begin{equation} \label{eq:solvabilityCondition_cg1mode}
\begin{pmatrix}
 \Delta m_0 - \frac{\partial U_0}{\partial T} \\
 \frac{1}{2}\Delta m_0 + \frac{\omega_{1}^{2}}{2}m_0^2 U_0 - \frac{\Pi_c}{2} m_0^3 - \frac{\partial m_0}{\partial T}
\end{pmatrix}_{\Omega t}
\cdot \begin{pmatrix}
\frac{1}{2}\frac{\omega_{1}^{2} + i\omega_{0}\sqrt{D}}{\omega_{1}^{2} + D} \\
1
\end{pmatrix} = 0,
\end{equation}
where $\Omega t$ denotes that the associated expression is restricted to terms oscillating at $\Omega t$. Performing the tedious algebra and the scalar product, we find:
\begin{equation} \label{eq:AmplEq_cg1mode}
 \frac{1}{\omega_{1}^{2}}\frac{dA}{dT} = A \frac{\Delta}{\omega_{1}^{2}} \frac{2 + \mu + i\sqrt{\mu}}{4} - A |A|^2 \frac{ \left( 1 + \mu \right)\left( 3\mu + i\sqrt{\mu} \right) }{2},
\end{equation}
where $\mu = D/\omega_{1}^{2}$. \\
\paragraph{Amplitude equation.} Finally, introducing $A = R e^{i\Psi}$, we obtain the amplitude equation for the real amplitude $R$:
\begin{equation} \label{eq:AmplitudeEquationR_cg1mode}
 \frac{1}{\omega_{1}^{2}}\frac{dR}{dT} = R \frac{\Delta}{\omega_{1}^{2}} \frac{2 + \mu}{4} - R^3 \frac{3\mu\left( 1+\mu \right)}{2}.
\end{equation}
At first order, we thus find that for $\Pi < \Pi_c$ ($\Delta < 0$), the only stable solution is the disordered fixed point; $A = 0$; and for $\Pi > \Pi_c$ ($\Delta > 0$), the only stable solution is the nonlinear limit cycle spontaneously oscillating along the lowest energy mode. As one approaches the bifurcation from above, the activity-independent frequency $\Omega = \sqrt{D\omega_{1}^{2}}$ remains the same, while the amplitude vanishes like a square root. The normal form, Eq. (\ref{eq:AmplitudeEquationR_cg1mode}), and the linear stability analysis are the hallmarks of a supercritical Hopf bifurcation. Importantly, in contrast with \textit{synchronized chiral oscillations}~\cite{baconnier2022selective}, at the level of homogeneous solutions of the coarse-grained model, we find a continuous transition from the disordered phase to NICA. \\
\paragraph{Timescales separation.} The long timescale corresponds to the typical relaxation timescale of the transitory regime $\tau_\textrm{relax} \simeq 1/\delta$; and the short one to the period of the oscillations $\tau_\textrm{osc} \simeq 1/\omega_{1}^{2}\sqrt{\mu}$. The timescale separation condition can be written as follows:
\begin{equation} \label{eq:timescale_separation}
 \delta \ll \omega_{0}\sqrt{D},
\end{equation}
which is verified close enough to the threshold.

\end{widetext}


\begin{thebibliography}{31}%
\makeatletter
\providecommand \@ifxundefined [1]{%
 \@ifx{#1\undefined}
}%
\providecommand \@ifnum [1]{%
 \ifnum #1\expandafter \@firstoftwo
 \else \expandafter \@secondoftwo
 \fi
}%
\providecommand \@ifx [1]{%
 \ifx #1\expandafter \@firstoftwo
 \else \expandafter \@secondoftwo
 \fi
}%
\providecommand \natexlab [1]{#1}%
\providecommand \enquote  [1]{``#1''}%
\providecommand \bibnamefont  [1]{#1}%
\providecommand \bibfnamefont [1]{#1}%
\providecommand \citenamefont [1]{#1}%
\providecommand \href@noop [0]{\@secondoftwo}%
\providecommand \href [0]{\begingroup \@sanitize@url \@href}%
\providecommand \@href[1]{\@@startlink{#1}\@@href}%
\providecommand \@@href[1]{\endgroup#1\@@endlink}%
\providecommand \@sanitize@url [0]{\catcode `\\12\catcode `\$12\catcode
  `\&12\catcode `\#12\catcode `\^12\catcode `\_12\catcode `\%12\relax}%
\providecommand \@@startlink[1]{}%
\providecommand \@@endlink[0]{}%
\providecommand \url  [0]{\begingroup\@sanitize@url \@url }%
\providecommand \@url [1]{\endgroup\@href {#1}{\urlprefix }}%
\providecommand \urlprefix  [0]{URL }%
\providecommand \Eprint [0]{\href }%
\providecommand \doibase [0]{http://dx.doi.org/}%
\providecommand \selectlanguage [0]{\@gobble}%
\providecommand \bibinfo  [0]{\@secondoftwo}%
\providecommand \bibfield  [0]{\@secondoftwo}%
\providecommand \translation [1]{[#1]}%
\providecommand \BibitemOpen [0]{}%
\providecommand \bibitemStop [0]{}%
\providecommand \bibitemNoStop [0]{.\EOS\space}%
\providecommand \EOS [0]{\spacefactor3000\relax}%
\providecommand \BibitemShut  [1]{\csname bibitem#1\endcsname}%
\let\auto@bib@innerbib\@empty
\bibitem [{\citenamefont {Woodhouse}\ \emph {et~al.}(2018)\citenamefont
  {Woodhouse}, \citenamefont {Ronellenfitsch},\ and\ \citenamefont
  {Dunkel}}]{woodhouse2018autonomous}%
  \BibitemOpen
  \bibfield  {author} {\bibinfo {author} {\bibfnamefont {F.~G.}\ \bibnamefont
  {Woodhouse}}, \bibinfo {author} {\bibfnamefont {H.}~\bibnamefont
  {Ronellenfitsch}}, \ and\ \bibinfo {author} {\bibfnamefont {J.}~\bibnamefont
  {Dunkel}},\ }\href@noop {} {\bibfield  {journal} {\bibinfo  {journal}
  {Physical Review Letters}\ }\textbf {\bibinfo {volume} {121}},\ \bibinfo
  {pages} {178001} (\bibinfo {year} {2018})}\BibitemShut {NoStop}%
\bibitem [{\citenamefont {Henkes}\ \emph {et~al.}(2011)\citenamefont {Henkes},
  \citenamefont {Fily},\ and\ \citenamefont {Marchetti}}]{henkes2011active}%
  \BibitemOpen
  \bibfield  {author} {\bibinfo {author} {\bibfnamefont {S.}~\bibnamefont
  {Henkes}}, \bibinfo {author} {\bibfnamefont {Y.}~\bibnamefont {Fily}}, \ and\
  \bibinfo {author} {\bibfnamefont {M.~C.}\ \bibnamefont {Marchetti}},\
  }\href@noop {} {\bibfield  {journal} {\bibinfo  {journal} {Physical Review
  E}\ }\textbf {\bibinfo {volume} {84}},\ \bibinfo {pages} {040301(R)} (\bibinfo
  {year} {2011})}\BibitemShut {NoStop}%
\bibitem [{\citenamefont {Ferrante}\ \emph {et~al.}(2013)\citenamefont
  {Ferrante}, \citenamefont {Turgut}, \citenamefont {Dorigo},\ and\
  \citenamefont {Huepe}}]{ferrante2013elasticity}%
  \BibitemOpen
  \bibfield  {author} {\bibinfo {author} {\bibfnamefont {E.}~\bibnamefont
  {Ferrante}}, \bibinfo {author} {\bibfnamefont {A.~E.}\ \bibnamefont
  {Turgut}}, \bibinfo {author} {\bibfnamefont {M.}~\bibnamefont {Dorigo}}, \
  and\ \bibinfo {author} {\bibfnamefont {C.}~\bibnamefont {Huepe}},\
  }\href@noop {} {\bibfield  {journal} {\bibinfo  {journal} {Physical review
  letters}\ }\textbf {\bibinfo {volume} {111}},\ \bibinfo {pages} {268302}
  (\bibinfo {year} {2013})}\BibitemShut {NoStop}%
\bibitem [{\citenamefont {Baconnier}\ \emph {et~al.}(2022)\citenamefont
  {Baconnier}, \citenamefont {Shohat}, \citenamefont {L{\'o}pez}, \citenamefont
  {Coulais}, \citenamefont {D{\'e}mery}, \citenamefont {D{\"u}ring},\ and\
  \citenamefont {Dauchot}}]{baconnier2022selective}%
  \BibitemOpen
  \bibfield  {author} {\bibinfo {author} {\bibfnamefont {P.}~\bibnamefont
  {Baconnier}}, \bibinfo {author} {\bibfnamefont {D.}~\bibnamefont {Shohat}},
  \bibinfo {author} {\bibfnamefont {C.~H.}\ \bibnamefont {L{\'o}pez}}, \bibinfo
  {author} {\bibfnamefont {C.}~\bibnamefont {Coulais}}, \bibinfo {author}
  {\bibfnamefont {V.}~\bibnamefont {D{\'e}mery}}, \bibinfo {author}
  {\bibfnamefont {G.}~\bibnamefont {D{\"u}ring}}, \ and\ \bibinfo {author}
  {\bibfnamefont {O.}~\bibnamefont {Dauchot}},\ }\href@noop {} {\bibfield
  {journal} {\bibinfo  {journal} {Nature Physics}\ }\textbf {\bibinfo {volume}
  {18}},\ \bibinfo {pages} {1234} (\bibinfo {year} {2022})}\BibitemShut
  {NoStop}%
\bibitem [{\citenamefont {Zheng}\ \emph {et~al.}(2022)\citenamefont {Zheng},
  \citenamefont {Huepe},\ and\ \citenamefont {Han}}]{zheng2022experimental}%
  \BibitemOpen
  \bibfield  {author} {\bibinfo {author} {\bibfnamefont {Y.}~\bibnamefont
  {Zheng}}, \bibinfo {author} {\bibfnamefont {C.}~\bibnamefont {Huepe}}, \ and\
  \bibinfo {author} {\bibfnamefont {Z.}~\bibnamefont {Han}},\ }\href@noop {}
  {\bibfield  {journal} {\bibinfo  {journal} {Adaptive Behavior}\ }\textbf
  {\bibinfo {volume} {30}},\ \bibinfo {pages} {19} (\bibinfo {year}
  {2022})}\BibitemShut {NoStop}%
\bibitem [{\citenamefont {Baconnier}\ \emph {et~al.}(2023)\citenamefont
  {Baconnier}, \citenamefont {Shohat},\ and\ \citenamefont
  {Dauchot}}]{baconnier2023discontinuous}%
  \BibitemOpen
  \bibfield  {author} {\bibinfo {author} {\bibfnamefont {P.}~\bibnamefont
  {Baconnier}}, \bibinfo {author} {\bibfnamefont {D.}~\bibnamefont {Shohat}}, \
  and\ \bibinfo {author} {\bibfnamefont {O.}~\bibnamefont {Dauchot}},\
  }\href@noop {} {\bibfield  {journal} {\bibinfo  {journal} {Physical Review
  Letters}\ }\textbf {\bibinfo {volume} {130}},\ \bibinfo {pages} {028201}
  (\bibinfo {year} {2023})}\BibitemShut {NoStop}%
\bibitem [{\citenamefont {Hern\'andez-L\'opez}\ \emph {et~al.}(2023)\citenamefont
  {Hern\'andez-L\'opez}, \citenamefont {Baconnier}, \citenamefont {Coulais},
  \citenamefont {Dauchot},\ and\ \citenamefont {Düring}}]{hernandez20232}%
  \BibitemOpen
  \bibfield  {author} {\bibinfo {author} {\bibfnamefont {C.}~\bibnamefont
  {Hern\'andez-L\'opez}}, \bibinfo {author} {\bibfnamefont {P.}~\bibnamefont
  {Baconnier}}, \bibinfo {author} {\bibfnamefont {C.}~\bibnamefont {Coulais}},
  \bibinfo {author} {\bibfnamefont {O.}~\bibnamefont {Dauchot}}, \ and\
  \bibinfo {author} {\bibfnamefont {G.}~\bibnamefont {Düring}},\ }\href
  {https:} {\bibfield  {journal} {\bibinfo  {journal} {arXiv:2310.12879}\ }
  (\bibinfo {year} {2023})}\BibitemShut {NoStop}%
\bibitem [{\citenamefont {Szabo}\ \emph {et~al.}(2006)\citenamefont {Szabo},
  \citenamefont {Sz{\"o}ll{\"o}si}, \citenamefont {G{\"o}nci}, \citenamefont
  {Jur{\'a}nyi}, \citenamefont {Selmeczi},\ and\ \citenamefont
  {Vicsek}}]{szabo2006phase}%
  \BibitemOpen
  \bibfield  {author} {\bibinfo {author} {\bibfnamefont {B.}~\bibnamefont
  {Szabo}}, \bibinfo {author} {\bibfnamefont {G.J.}~\bibnamefont
  {Sz{\"o}ll{\"o}si}}, \bibinfo {author} {\bibfnamefont {B.}~\bibnamefont
  {G{\"o}nci}}, \bibinfo {author} {\bibfnamefont {Z.}~\bibnamefont
  {Jur{\'a}nyi}}, \bibinfo {author} {\bibfnamefont {D.}~\bibnamefont
  {Selmeczi}}, \ and\ \bibinfo {author} {\bibfnamefont {T.}~\bibnamefont
  {Vicsek}},\ }\href@noop {} {\bibfield  {journal} {\bibinfo  {journal}
  {Physical Review E}\ }\textbf {\bibinfo {volume} {74}},\ \bibinfo {pages}
  {061908} (\bibinfo {year} {2006})}\BibitemShut {NoStop}%
\bibitem [{\citenamefont {Serra-Picamal}\ \emph {et~al.}(2012)\citenamefont
  {Serra-Picamal}, \citenamefont {Conte}, \citenamefont {Vincent},
  \citenamefont {Anon}, \citenamefont {Tambe}, \citenamefont {Bazellieres},
  \citenamefont {Butler}, \citenamefont {Fredberg},\ and\ \citenamefont
  {Trepat}}]{serra2012mechanical}%
  \BibitemOpen
  \bibfield  {author} {\bibinfo {author} {\bibfnamefont {X.}~\bibnamefont
  {Serra-Picamal}}, \bibinfo {author} {\bibfnamefont {V.}~\bibnamefont
  {Conte}}, \bibinfo {author} {\bibfnamefont {R.}~\bibnamefont {Vincent}},
  \bibinfo {author} {\bibfnamefont {E.}~\bibnamefont {Anon}}, \bibinfo {author}
  {\bibfnamefont {D.~T.}\ \bibnamefont {Tambe}}, \bibinfo {author}
  {\bibfnamefont {E.}~\bibnamefont {Bazellieres}}, \bibinfo {author}
  {\bibfnamefont {J.~P.}\ \bibnamefont {Butler}}, \bibinfo {author}
  {\bibfnamefont {J.~J.}\ \bibnamefont {Fredberg}}, \ and\ \bibinfo {author}
  {\bibfnamefont {X.}~\bibnamefont {Trepat}},\ }\href@noop {} {\bibfield
  {journal} {\bibinfo  {journal} {Nature Physics}\ }\textbf {\bibinfo {volume}
  {8}},\ \bibinfo {pages} {628} (\bibinfo {year} {2012})}\BibitemShut {NoStop}%
\bibitem [{\citenamefont {Deforet}\ \emph {et~al.}(2014)\citenamefont
  {Deforet}, \citenamefont {Hakim}, \citenamefont {Yevick}, \citenamefont
  {Duclos},\ and\ \citenamefont {Silberzan}}]{deforet2014emergence}%
  \BibitemOpen
  \bibfield  {author} {\bibinfo {author} {\bibfnamefont {M.}~\bibnamefont
  {Deforet}}, \bibinfo {author} {\bibfnamefont {V.}~\bibnamefont {Hakim}},
  \bibinfo {author} {\bibfnamefont {H.~G.}\ \bibnamefont {Yevick}}, \bibinfo
  {author} {\bibfnamefont {G.}~\bibnamefont {Duclos}}, \ and\ \bibinfo {author}
  {\bibfnamefont {P.}~\bibnamefont {Silberzan}},\ }\href@noop {} {\bibfield
  {journal} {\bibinfo  {journal} {Nature communications}\ }\textbf {\bibinfo
  {volume} {5}},\ \bibinfo {pages} {3747} (\bibinfo {year} {2014})}\BibitemShut
  {NoStop}%
\bibitem [{\citenamefont {Barton}\ \emph {et~al.}(2017)\citenamefont {Barton},
  \citenamefont {Henkes}, \citenamefont {Weijer},\ and\ \citenamefont
  {Sknepnek}}]{barton2017active}%
  \BibitemOpen
  \bibfield  {author} {\bibinfo {author} {\bibfnamefont {D.~L.}\ \bibnamefont
  {Barton}}, \bibinfo {author} {\bibfnamefont {S.}~\bibnamefont {Henkes}},
  \bibinfo {author} {\bibfnamefont {C.~J.}\ \bibnamefont {Weijer}}, \ and\
  \bibinfo {author} {\bibfnamefont {R.}~\bibnamefont {Sknepnek}},\ }\href@noop
  {} {\bibfield  {journal} {\bibinfo  {journal} {PLoS computational biology}\
  }\textbf {\bibinfo {volume} {13}},\ \bibinfo {pages} {e1005569} (\bibinfo
  {year} {2017})}\BibitemShut {NoStop}%
\bibitem [{\citenamefont {Petrolli}\ \emph {et~al.}(2019)\citenamefont
  {Petrolli}, \citenamefont {Le~Goff}, \citenamefont {Tadrous}, \citenamefont
  {Martens}, \citenamefont {Allier}, \citenamefont {Mandula}, \citenamefont
  {Herv{\'e}}, \citenamefont {Henkes}, \citenamefont {Sknepnek}, \citenamefont
  {Boudou} \emph {et~al.}}]{petrolli2019confinement}%
  \BibitemOpen
  \bibfield  {author} {\bibinfo {author} {\bibfnamefont {V.}~\bibnamefont
  {Petrolli}}, \bibinfo {author} {\bibfnamefont {M.}~\bibnamefont {Le~Goff}},
  \bibinfo {author} {\bibfnamefont {M.}~\bibnamefont {Tadrous}}, \bibinfo
  {author} {\bibfnamefont {K.}~\bibnamefont {Martens}}, \bibinfo {author}
  {\bibfnamefont {C.}~\bibnamefont {Allier}}, \bibinfo {author} {\bibfnamefont
  {O.}~\bibnamefont {Mandula}}, \bibinfo {author} {\bibfnamefont
  {L.}~\bibnamefont {Herv{\'e}}}, \bibinfo {author} {\bibfnamefont
  {S.}~\bibnamefont {Henkes}}, \bibinfo {author} {\bibfnamefont
  {R.}~\bibnamefont {Sknepnek}}, \bibinfo {author} {\bibfnamefont
  {T.}~\bibnamefont {Boudou}},  \emph {et~al.},\ }\href@noop {} {\bibfield
  {journal} {\bibinfo  {journal} {Physical review letters}\ }\textbf {\bibinfo
  {volume} {122}},\ \bibinfo {pages} {168101} (\bibinfo {year}
  {2019})}\BibitemShut {NoStop}%
\bibitem [{\citenamefont {Peyret}\ \emph {et~al.}(2019)\citenamefont {Peyret},
  \citenamefont {Mueller}, \citenamefont {d’Alessandro}, \citenamefont
  {Begnaud}, \citenamefont {Marcq}, \citenamefont {M{\`e}ge}, \citenamefont
  {Yeomans}, \citenamefont {Doostmohammadi},\ and\ \citenamefont
  {Ladoux}}]{peyret2019sustained}%
  \BibitemOpen
  \bibfield  {author} {\bibinfo {author} {\bibfnamefont {G.}~\bibnamefont
  {Peyret}}, \bibinfo {author} {\bibfnamefont {R.}~\bibnamefont {Mueller}},
  \bibinfo {author} {\bibfnamefont {J.}~\bibnamefont {d’Alessandro}},
  \bibinfo {author} {\bibfnamefont {S.}~\bibnamefont {Begnaud}}, \bibinfo
  {author} {\bibfnamefont {P.}~\bibnamefont {Marcq}}, \bibinfo {author}
  {\bibfnamefont {R.-M.}\ \bibnamefont {M{\`e}ge}}, \bibinfo {author}
  {\bibfnamefont {J.~M.}\ \bibnamefont {Yeomans}}, \bibinfo {author}
  {\bibfnamefont {A.}~\bibnamefont {Doostmohammadi}}, \ and\ \bibinfo {author}
  {\bibfnamefont {B.}~\bibnamefont {Ladoux}},\ }\href@noop {} {\bibfield
  {journal} {\bibinfo  {journal} {Biophysical journal}\ }\textbf {\bibinfo
  {volume} {117}},\ \bibinfo {pages} {464} (\bibinfo {year}
  {2019})}\BibitemShut {NoStop}%
\bibitem [{\citenamefont {Liu}\ \emph {et~al.}(2021)\citenamefont {Liu},
  \citenamefont {Shankar}, \citenamefont {Marchetti},\ and\ \citenamefont
  {Wu}}]{liu2021viscoelastic}%
  \BibitemOpen
  \bibfield  {author} {\bibinfo {author} {\bibfnamefont {S.}~\bibnamefont
  {Liu}}, \bibinfo {author} {\bibfnamefont {S.}~\bibnamefont {Shankar}},
  \bibinfo {author} {\bibfnamefont {M.~C.}\ \bibnamefont {Marchetti}}, \ and\
  \bibinfo {author} {\bibfnamefont {Y.}~\bibnamefont {Wu}},\ }\href@noop {}
  {\bibfield  {journal} {\bibinfo  {journal} {Nature}\ }\textbf {\bibinfo
  {volume} {590}},\ \bibinfo {pages} {80} (\bibinfo {year} {2021})}\BibitemShut
  {NoStop}%
\bibitem [{\citenamefont {Xu}\ \emph {et~al.}(2023)\citenamefont {Xu},
  \citenamefont {Huang}, \citenamefont {Zhang},\ and\ \citenamefont
  {Wu}}]{xu2023autonomous}%
  \BibitemOpen
  \bibfield  {author} {\bibinfo {author} {\bibfnamefont {H.}~\bibnamefont
  {Xu}}, \bibinfo {author} {\bibfnamefont {Y.}~\bibnamefont {Huang}}, \bibinfo
  {author} {\bibfnamefont {R.}~\bibnamefont {Zhang}}, \ and\ \bibinfo {author}
  {\bibfnamefont {Y.}~\bibnamefont {Wu}},\ }\href@noop {} {\bibfield  {journal}
  {\bibinfo  {journal} {Nature Physics}\ }\textbf {\bibinfo {volume} {19}},\
  \bibinfo {pages} {46} (\bibinfo {year} {2023})}\BibitemShut {NoStop}%
\bibitem [{\citenamefont {Dauchot}\ and\ \citenamefont
  {D{\'e}mery}(2019)}]{dauchot2019dynamics}%
  \BibitemOpen
  \bibfield  {author} {\bibinfo {author} {\bibfnamefont {O.}~\bibnamefont
  {Dauchot}}\ and\ \bibinfo {author} {\bibfnamefont {V.}~\bibnamefont
  {D{\'e}mery}},\ }\href@noop {} {\bibfield  {journal} {\bibinfo  {journal}
  {Physical review letters}\ }\textbf {\bibinfo {volume} {122}},\ \bibinfo
  {pages} {068002} (\bibinfo {year} {2019})}\BibitemShut {NoStop}%
\bibitem [{\citenamefont {Damascena}\ \emph {et~al.}(2022)\citenamefont
  {Damascena}, \citenamefont {Cabral},\ and\ \citenamefont
  {de~Souza~Silva}}]{damascena2022coexisting}%
  \BibitemOpen
  \bibfield  {author} {\bibinfo {author} {\bibfnamefont {R.~H.}\ \bibnamefont
  {Damascena}}, \bibinfo {author} {\bibfnamefont {L.~R.~E.}\ \bibnamefont
  {Cabral}}, \ and\ \bibinfo {author} {\bibfnamefont {C.~C.}\ \bibnamefont
  {de~Souza~Silva}},\ }\href@noop {} {\bibfield  {journal} {\bibinfo  {journal}
  {Physical Review E}\ }\textbf {\bibinfo {volume} {105}},\ \bibinfo {pages}
  {064608} (\bibinfo {year} {2022})}\BibitemShut {NoStop}%
\bibitem [{sup()}]{supplentary_information}%
  \BibitemOpen
  \href@noop {} {\ See Supplemental Material at [url] for movies and normal mode spectrums}\BibitemShut {NoStop}%
\bibitem [{\citenamefont {Jakobsen}(2013)}]{jakobsen2013introduction}%
  \BibitemOpen
  \bibfield  {author} {\bibinfo {author} {\bibfnamefont {P.}~\bibnamefont
  {Jakobsen}},\ }\href@noop {} {\bibfield  {journal} {\bibinfo {journal}
  {arXiv:1312.3651}\ } (\bibinfo {year} {2013})}\BibitemShut
  {NoStop}%
\bibitem [{\citenamefont {Kuramoto}(1975)}]{kuramoto1975self}%
  \BibitemOpen
  \bibfield  {author} {\bibinfo {author} {\bibfnamefont {Y.}~\bibnamefont
  {Kuramoto}},\ }in\ \href@noop {} {\emph {\bibinfo {booktitle} {International
  Symposium on Mathematical Problems in Theoretical Physics: January 23--29,
  1975, Kyoto University, Kyoto, Japan}}}\ (\bibinfo {organization} {Springer},\
  \bibinfo {year} {1975})\ pp.\ \bibinfo {pages} {420--422}\BibitemShut
  {NoStop}%
\bibitem [{\citenamefont {Mirollo}\ and\ \citenamefont
  {Strogatz}(1990)}]{mirollo1990synchronization}%
  \BibitemOpen
  \bibfield  {author} {\bibinfo {author} {\bibfnamefont {R.~E.}\ \bibnamefont
  {Mirollo}}\ and\ \bibinfo {author} {\bibfnamefont {S.~H.}\ \bibnamefont
  {Strogatz}},\ }\href@noop {} {\bibfield  {journal} {\bibinfo  {journal} {SIAM
  Journal on Applied Mathematics}\ }\textbf {\bibinfo {volume} {50}},\ \bibinfo
  {pages} {1645} (\bibinfo {year} {1990})}\BibitemShut {NoStop}%
\bibitem [{\citenamefont {Strogatz}\ and\ \citenamefont
  {Stewart}(1993)}]{strogatz1993coupled}%
  \BibitemOpen
  \bibfield  {author} {\bibinfo {author} {\bibfnamefont {S.~H.}\ \bibnamefont
  {Strogatz}}\ and\ \bibinfo {author} {\bibfnamefont {I.}~\bibnamefont
  {Stewart}},\ }\href@noop {} {\bibfield  {journal} {\bibinfo  {journal}
  {Scientific american}\ }\textbf {\bibinfo {volume} {269}},\ \bibinfo {pages}
  {102} (\bibinfo {year} {1993})}\BibitemShut {NoStop}%
\bibitem [{\citenamefont {Pikovsky}\ \emph {et~al.}(2001)\citenamefont
  {Pikovsky}, \citenamefont {Rosenblum},\ and\ \citenamefont {Kurths}}]{pikovsky2001universal}%
  \BibitemOpen
  \bibfield  {author} {\bibinfo {author} {\bibfnamefont {A.}~\bibnamefont
  {Pikovsky}}, \bibinfo {author} {\bibfnamefont {M.}~\bibnamefont {Rosenblum}}, \and\
  \bibinfo {author} {\bibfnamefont {J.}~\bibnamefont {Kurths}},\ }\href@noop {}
  {\bibfield  {journal} {\bibinfo  {journal} {Cambridge Non-linear Science Series (Cambridge University Press)}\ } (\bibinfo {year} {2001})}\BibitemShut
  {NoStop}%
\bibitem [{\citenamefont {Strogatz}(2004)}]{strogatz2004sync}%
  \BibitemOpen
  \bibfield  {author} {\bibinfo {author} {\bibfnamefont {S.}~\bibnamefont
  {Strogatz}},\ }\href@noop {} {\bibinfo {year} {2004},\ }\bibinfo {publisher} {Penguin Press Science}\BibitemShut {NoStop}%
\bibitem [{\citenamefont {Acebr{\'o}n}\ \emph {et~al.}(2005)\citenamefont
  {Acebr{\'o}n}, \citenamefont {Bonilla}, \citenamefont {Vicente},
  \citenamefont {Ritort},\ and\ \citenamefont {Spigler}}]{acebron2005kuramoto}%
  \BibitemOpen
  \bibfield  {author} {\bibinfo {author} {\bibfnamefont {J.~A.}\ \bibnamefont
  {Acebr{\'o}n}}, \bibinfo {author} {\bibfnamefont {L.~L.}\ \bibnamefont
  {Bonilla}}, \bibinfo {author} {\bibfnamefont {C.~J.~P.}\ \bibnamefont
  {Vicente}}, \bibinfo {author} {\bibfnamefont {F.}~\bibnamefont {Ritort}}, \
  and\ \bibinfo {author} {\bibfnamefont {R.}~\bibnamefont {Spigler}},\
  }\href@noop {} {\bibfield  {journal} {\bibinfo  {journal} {Reviews of modern
  physics}\ }\textbf {\bibinfo {volume} {77}},\ \bibinfo {pages} {137}
  (\bibinfo {year} {2005})}\BibitemShut {NoStop}%
\bibitem [{\citenamefont {O'Keeffe}\ \emph {et~al.}(2017)\citenamefont
  {O'Keeffe}, \citenamefont {Hong},\ and\ \citenamefont
  {Strogatz}}]{o2017oscillators}%
  \BibitemOpen
  \bibfield  {author} {\bibinfo {author} {\bibfnamefont {K.~P.}\ \bibnamefont
  {O'Keeffe}}, \bibinfo {author} {\bibfnamefont {H.}~\bibnamefont {Hong}}, \
  and\ \bibinfo {author} {\bibfnamefont {S.~H.}\ \bibnamefont {Strogatz}},\
  }\href@noop {} {\bibfield  {journal} {\bibinfo  {journal} {Nature
  communications}\ }\textbf {\bibinfo {volume} {8}},\ \bibinfo {pages} {1504}
  (\bibinfo {year} {2017})}\BibitemShut {NoStop}%
\bibitem [{\citenamefont {Zheng}\ \emph {et~al.}(2023)\citenamefont {Zheng},
  \citenamefont {Brandenbourger}, \citenamefont {Robinet}, \citenamefont
  {Schall}, \citenamefont {Lerner},\ and\ \citenamefont
  {Coulais}}]{zheng2023self}%
  \BibitemOpen
  \bibfield  {author} {\bibinfo {author} {\bibfnamefont {E.}~\bibnamefont
  {Zheng}}, \bibinfo {author} {\bibfnamefont {M.}~\bibnamefont
  {Brandenbourger}}, \bibinfo {author} {\bibfnamefont {L.}~\bibnamefont
  {Robinet}}, \bibinfo {author} {\bibfnamefont {P.}~\bibnamefont {Schall}},
  \bibinfo {author} {\bibfnamefont {E.}~\bibnamefont {Lerner}}, \ and\ \bibinfo
  {author} {\bibfnamefont {C.}~\bibnamefont {Coulais}},\ }\href@noop {}
  {\bibfield  {journal} {\bibinfo  {journal} {Physical Review Letters}\
  }\textbf {\bibinfo {volume} {130}},\ \bibinfo {pages} {178202} (\bibinfo
  {year} {2023})}\BibitemShut {NoStop}%
\bibitem [{\citenamefont {Henkes}\ \emph {et~al.}(2012)\citenamefont {Henkes},
  \citenamefont {Brito},\ and\ \citenamefont {Dauchot}}]{henkes2012extracting}%
  \BibitemOpen
  \bibfield  {author} {\bibinfo {author} {\bibfnamefont {S.}~\bibnamefont
  {Henkes}}, \bibinfo {author} {\bibfnamefont {C.}~\bibnamefont {Brito}}, \
  and\ \bibinfo {author} {\bibfnamefont {O.}~\bibnamefont {Dauchot}},\
  }\href@noop {} {\bibfield  {journal} {\bibinfo  {journal} {Soft Matter}\
  }\textbf {\bibinfo {volume} {8}},\ \bibinfo {pages} {6092} (\bibinfo {year}
  {2012})}\BibitemShut {NoStop}%
\bibitem [{\citenamefont {Dotsenko}(1995)}]{dotsenko1995critical}%
  \BibitemOpen
  \bibfield  {author} {\bibinfo {author} {\bibfnamefont {V.~S.}\ \bibnamefont
  {Dotsenko}},\ }\href@noop {} {\bibfield  {journal} {\bibinfo  {journal}
  {Physics-Uspekhi}\ }\textbf {\bibinfo {volume} {38}},\ \bibinfo {pages} {457}
  (\bibinfo {year} {1995})}\BibitemShut {NoStop}%
\bibitem [{\citenamefont {Bechinger}\ \emph {et~al.}(2016)\citenamefont
  {Bechinger}, \citenamefont {Di~Leonardo}, \citenamefont {L{\"o}wen},
  \citenamefont {Reichhardt}, \citenamefont {Volpe},\ and\ \citenamefont
  {Volpe}}]{bechinger2016active}%
  \BibitemOpen
  \bibfield  {author} {\bibinfo {author} {\bibfnamefont {C.}~\bibnamefont
  {Bechinger}}, \bibinfo {author} {\bibfnamefont {R.}~\bibnamefont
  {Di~Leonardo}}, \bibinfo {author} {\bibfnamefont {H.}~\bibnamefont
  {L{\"o}wen}}, \bibinfo {author} {\bibfnamefont {C.}~\bibnamefont
  {Reichhardt}}, \bibinfo {author} {\bibfnamefont {G.}~\bibnamefont {Volpe}}, \
  and\ \bibinfo {author} {\bibfnamefont {G.}~\bibnamefont {Volpe}},\
  }\href@noop {} {\bibfield  {journal} {\bibinfo  {journal} {Reviews of Modern
  Physics}\ }\textbf {\bibinfo {volume} {88}},\ \bibinfo {pages} {045006}
  (\bibinfo {year} {2016})}\BibitemShut {NoStop}%
\bibitem [{\citenamefont {Risken}(1996)}]{Risken1996}%
  \BibitemOpen
  \bibfield  {author} {\bibinfo {author} {\bibfnamefont {H.}~\bibnamefont
  {Risken}},\ }\href
  {{https://link.springer.com/book/10.1007/978-3-642-61544-3}} {\emph {\bibinfo
  {title} {{The Fokker-Planck Equation: Methods of Solutions and
  Applications}}}},\ \bibinfo {edition} {{2nd ed.}},\
  {Springer Series in Synergetics}\ (\bibinfo  {publisher} {{Springer}},\
  \bibinfo {year} {1996})\BibitemShut {NoStop}%
\end{thebibliography}
\end{document}